\documentclass[preprint,12pt]{elsarticle}
 \usepackage{amsmath}
\usepackage{hyperref} 

\journal{Journal of \LaTeX\ Templates}

\biboptions{sort&compress}
\begin{document}

\begin{frontmatter}

\title{Feasibility of radar detection of extensive air
showers} 


\author[ifj]{J. Stasielak}
\ead{jaroslaw.stasielak@ifj.edu.pl}

\author[KIT]{R. Engel}
\author[KIT]{S. Baur}
\author[bonn]{P. Neunteufel}
\author[KIT]{R. \v{S}m\'{\i}da}
\author[KIT]{F. Werner\corref{now}}
\cortext[now]{now at Max Planck Institute for Nuclear Physics, P.O. Box 103980, D 69029 Heidelberg, Germany}
\author[ifj]{H. Wilczy\'nski}

\address[ifj]{Institute of Nuclear Physics PAN, Krak\'ow, Poland}
\address[KIT]{Institute for Nuclear Physics (IKP), Karlsruhe Institute of Technology, Karlsruhe, Germany}
\address[bonn]{Argelander Institute for Astronomy (AIfA), Bonn, Germany}

\begin{abstract}
Reflection of radio waves off the short-lived plasma produced by the high-energy shower particles in the air is simulated, considering various radar setups and shower geometries. We show that the plasma produced by air showers has to be treated always as underdense. Therefore, we use the Thomson cross-section for scattering of  radio waves corrected for molecular quenching and we sum coherently contributions of the reflected radio wave over the volume of the plasma disk to obtain the time evolution of the signal arriving at the receiver antenna. The received power and the spectral power density of the radar echo are analyzed. Based on the obtained results, we discuss possible modes of radar detection of extensive air showers. We conclude that the scattered signal is too weak for the radar method to provide an efficient and inexpensive method of air shower detection.
\end{abstract}


\end{frontmatter}


\section{Introduction}

Traditional techniques of extensive air shower detection include recording the shower particles at the ground level or optical methods such as measuring fluorescence light from nitrogen in the atmosphere excited by the shower particles and Cherenkov light of air showers. Detecting radio emission from the shower particles in the MHz--GHz frequency range is a promising new technique that is currently studied, since it offers the possibility of about $100\%$ duty cycle in comparison to only about $15\%$ for optical methods \cite{bibe:LOPES,bibe:CODALEMA,bibe:AERA,bibe:MIDAS,bibe:AMBER,bibe:EASIER,bibe:CROME,bibe:smida}. An additional method is the radar technique, in which a ground-based radio transmitter irradiates the ionization trail left behind the shower front and another ground-based antenna receives the scattered radio signal. This remote sensing technology could allow the construction of cosmic ray observatories with very large apertures to be built at much lower cost, with almost 100$\%$ duty cycle. 

The concept of implementing a radar technique for cosmic ray detection dates back to 1940 \cite{bibe:blackett}. However, due to the lack of experimental confirmation of shower signals, this method was not pursued for several decades. In recent years, renewed attention has been given to this topic \cite{bibe:baruch,bibe:gorham,bibe:gorham2,bibe:bakunov,bibe:bakunov2,bibe:takai,bibe:stasielak,bibe:stasielak2,bibe:filonenko,bibe:vries} and experimental efforts to detect cosmic ray showers using the radar technique were made by several groups \cite{bibe:matano,bibe:vin,bibe:lyono,wahl,bibe:terasawa,bibe:mariachi1,bibe:mariachi2,bibe:tara,bibe:tara2,bibe:tara3,bibe:ikeda,bibe:tara4}.

The radar technique has been used already for decades to observe meteors and lightnings. The detection method is based on the principle of scattering of radio waves off the plasma produced in the atmosphere by the passing meteor or lightning discharge. After the ionization trail is formed, the free electron concentration decreases because of diffusion, atmospheric turbulence, and loss of ionization through recombination and ionic reactions. Among these effects the most important one for meteors is  diffusion, which reduces the volume density, but essentially leaves the line density unchanged due to the long ionization trails of  meteors.

Meteors are observed at altitudes of 80 to 120 km and their typical velocities are in the range of 10 to 70 km/s. The ionization trails produced by them have long lifetimes, thus they can extend over several kilometers and have an initial radius of order of 1 m up to even 10 m. The ionization column of a meteor has an approximately uniform radial distribution of density and a line density of electrons between $10^{11}$ and $10^{14}$ cm$^{-1}$ \cite{meteor}. The plasma electrons can be considered to be in thermal equilibrium with the ambient atmosphere except for the very early stages of the formation of the meteor trail. Densities of the ionization trails produced by lightnings are many orders of magnitude higher than those produced by meteors and the plasma channels are of much smaller sizes.

The ionization trail that results from meteors or lightnings is traditionally divided  into the underdense and overdense regions, depending on the local characteristic plasma frequency $\omega_p$. If the electron density is high enough that the plasma frequency exceeds the radar frequency, i.e. the frequency of the emitted radio wave, then the radio wave is reflected from its surface. Such a region is called overdense. In contrast, if the electron density is low enough that the local plasma frequency is lower than the frequency of the incoming radio wave, then the region is underdense and the radio wave can penetrate the ionized region. In such a case the reflections by scattering of the radio wave off individual free electrons has to be considered. For radar frequencies used in meteor science, the limiting density between these two regimes is around the ionization line density of $10^{12}$ cm$^{-1}$.

Gorham \cite{bibe:gorham} considered radar reflection from the side of a horizontal ionization trail left by an air shower caused by ultra-high energy neutrinos at an altitude of about 10 km. He suggested that the most inner (densest) part of the ionization column is responsible for the bulk of the radar reflection. In analogy with the reflective behaviour of the overdense region produced by a meteor, he assumed that the radar cross-section of the overdense trail produced by a shower corresponds to the radar cross-section of a thin metallic wire. 

An alternative mode of shower detection was discussed in \cite{bibe:bakunov,bibe:bakunov2}, where reflection of the radar wave from the relativistically moving shower front was considered. The reflection coefficient was obtained by solving the Maxwell equations with the corresponding boundary conditions. In this mode of detection, the frequency of the radar echo is higher than the frequency of the incoming wave, in contrast to the reflection from the side of the ionization trail, where the frequency change is very small.

Scattering of a radio wave from the ionization trail produced by a shower in the underdense plasma regime was considered in \cite{bibe:takai}. The calculations were made for the forward scattered signal assuming that the ionization occurs in a line along the shower axis, i.e. that contributions from the laterally distributed electrons are coherent. The transmitter and receiver were located 50 km apart.

Finally, Filonenko \cite{bibe:filonenko} calculated the signal reflected from the plasma disk produced at the shower maximum at altitude of about 4 km.  He solved the equation of motion of plasma electrons in the electromagnetic field, accounting for electron collisions with neutral molecules.

The ionization trail that is produced in the  atmosphere by the passage of the high-energy particles of a shower (shower front), consists of electrons, which are essentially at rest with respect to the surrounding atmosphere. The plasma decays exponentially with the lifetime that depends on the air density and is equal to 15 ns at sea level \cite{bibe:vidmar,bibe:nijdam}, whereas the radial dependence of its density is controlled by the lateral distribution of the shower energy deposition in the air. The electron density is highest at the shower axis and decreases steeply with the distance from it. The diameter of the ionization trail is of several hundred meters. The shower front moves approximately with the speed of light in vacuum. Due to the short lifetime of the created plasma, the plasma-filled region behind the shower front also moves with the speed of light even though the electrons of the plasma do not move on macroscopic scales. Therefore, a Doppler effect will be observed in the radar echo, unless the shower is seen from the side. An enhancement of the signal scattered backwards due to its time compression is also expected. 

In this paper, which is an extension of our previous work \cite{bibe:stasielak,bibe:stasielak2}, we investigate the feasibility of detecting extensive air showers by the radar technique. Simulations are performed for the underdense regime using the Thomson cross-section for scattering of radio waves off the short-lived, non-moving plasma, with a correction for molecular quenching. We sum coherently the reflected radio waves off the individual electrons over the volume of the disk-like ionization trail and obtain the time evolution of the radar echo.

As an application we will consider the feasibility of supplementing a microwave detector \cite{bibe:MIDAS,bibe:AMBER,bibe:EASIER,bibe:CROME} with a radio transmitter and using the microwave antennas (tuned to the GHz frequency band) as receivers for the radar echo.
Furthermore, we will check whether the CROME (Cosmic Ray Observation via Microwave Emission) results \cite{bibe:CROME,bibe:smida,bibe:werner} could be interpreted as the radar reflection from the plasma produced by the shower in the air.

The outline of the paper is as follows. In Section \ref{modeling} we describe the details of our calculations. Section \ref{air-plasma} contains an extensive discussion of the plasma properties. In Section \ref{SimAnalyze} we describe the simulations performed  and present the analysis of the radar echo. The results and accuracy of the current approach to the problem of the air shower detection by radar is discussed in Section \ref{diss}. Finally, the Appendix contains the detailed derivation of the formulae from Section \ref{modeling}.

\section{Modeling radar reflection}

\label{modeling}

\subsection{Calculation of the radar signal}

 \begin{figure}[t]
  \centering
  \includegraphics[width=0.7\textwidth]{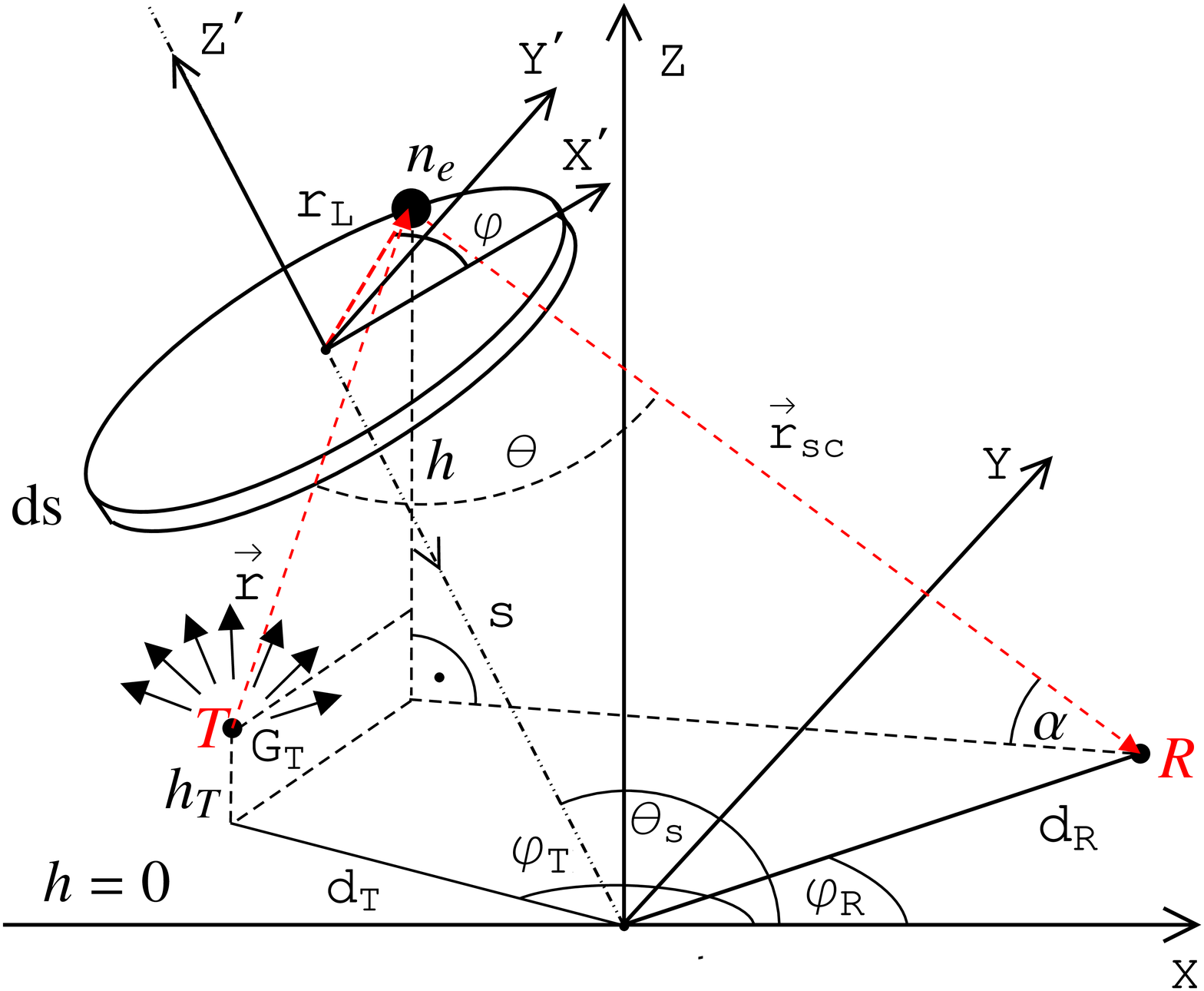}
  \caption{Schematic diagram representing the considered radar system and reflection from the plasma disk produced by a shower in the atmosphere. A ground-based radio transmitter (T) emits a radio signal, which is scattered off an element of the plasma disk and subsequently observed by the receiver antenna (R). The geometry of the radar system is determined by the distances from the shower core to the transmitter ($d_T$) and to the receiver ($d_R$) together with their azimuth angles ($\varphi_T$ and $\varphi_R$) and the altitude of the transmitter ($h_T$).}
  \label{fig:1}
 \end{figure}
 
A schematic diagram representing the concept of extensive air shower detection using the radar technique is shown in Figure \ref{fig:1}.
A ground-based radio transmitter ($T$) irradiates a disk-like static plasma left behind the shower front. The radio signal is scattered by free electrons in the ionization trail and subsequently received by the ground-based antenna ($R$). The geometry of such a radar system is described conveniently by the cylindrical coordinates of the transmitter and the receiver, i.e. by the distances from the shower core to the transmitter ($d_T$) and to the receiver ($d_R$), and by the angles $\varphi_T$ and $\varphi_R$. The altitude of the transmitter is given by $h_T$, whereas the receiver is at the ground level.

The system of coordinates $XYZ$ is chosen in such a way that the plane constructed by the $X$-axis and the shower axis is perpendicular to the ground. Moreover, the $X$ and $Y$ axes lie at the ground level. The center of the coordinate system is placed at the shower core, i.e. at the point of intersection of the shower axis with the ground. The coordinate system of the disk-like static plasma $X'Y'Z'$ is simply created by rotating the $XYZ$ frame of reference around $Y$-axis by the shower inclination angle $\left| \theta_s-\pi/2 \right|$
 and translating the resulting system of coordinates along $Z'$-axis by $s$.

The disk in Figure \ref{fig:1}, which is perpendicular to the shower axis, represents a slice of the static plasma. Its distance to the shower core at ground is equal to $s$. In the following we will consider a plasma volume element ${\rm d}V$ with polar coordinates  $r_L$ (distance from the shower axis) and azimuth $\varphi$.

 The electric field of the continuously emitted, incoming radio wave in this volume element at a given time $t_{r}$ can be written as
\begin{equation}
U_{\rm{inc}}  ~=~ U_{T} \sqrt{G_{T}} e^{i(\omega t_{r} + \phi_0)} 
\frac{e^{-i \int_{{\bf r}} n \hspace{0.1cm} {\bf k} \cdot  {\rm d}{\bf r}} }{ |{\bf r}|} \label{Uinc} \rm{,}
\end{equation}
where $\bf k$ ($	| \textbf{k} | =2\pi/\lambda$) is the wavevector of the emitted wave, $\lambda$ and $\omega$ are the wavelength and radian frequency of the emitted radio wave, $U_{T}$ is the amplitude of the transmitted field, $G_{T}$ is the
transmitter gain factor ($G_{T}=1$ for isotropic emission), and $\phi_0$ is the initial phase of emitted signal. Here $n$ denotes the refractive index of the air:
\begin{equation}
	n =1 + 2.8 \times 10^{-4} \rho(h) /\rho (0) \rm{,}
\end{equation}
where $\rho(h)$ is the density of air at an altitude $h$ derived from the U.S. Standard Atmosphere model \cite{bibe:USstandard}.

The strength of the radio wave reflected from the plasma volume ${\rm d}V$ that arrives at the receiver antenna can be expressed by 
\begin{equation}
{\rm d}U_{\rm{rcv}} ~=~  U_{\rm{inc}} 
 e^{i\omega (t-t_{r})} e^{-i\int_{{\bf r}_{\rm{sc}}} n \hspace{0.1cm}{\bf k}_{\rm{sc}} \cdot {\rm d}{\bf r}_{\rm{sc}}} 
 \sqrt{\left(\frac{\omega}{\nu_c}\right)^2 \frac{{\rm d} \sigma_T }{{\rm d} \Omega} \Delta \Omega_{\rm{sc}}} n_e  {\rm d}V \rm{,} \label{dUrcv}
\end{equation}
where ${\bf k}_{\rm{sc}}$, $t$, and $\Delta \Omega_{\rm{sc}}$
are the wavevector of the scattered wave, the  arrival time of the signal to the detector, and the solid angle of the receiver as seen from the point of scattering, respectively. The angle-dependent Thomson cross-section ${\rm d} \sigma_T/{\rm d} \Omega$ is given by
\begin{equation}
{\rm d} \sigma_T/{\rm d} \Omega =  \frac{3}{16 \pi} \sigma_T  (1+\cos^2 \theta) \rm{,}	
\end{equation}
where $\sigma_T$ and $\theta$ are the total Thomson cross-section and the angle between the directions of incident and scattered radio waves. 
The factor $(\omega / \nu_c)^2$ is a correction for molecular quenching (see Section \ref{sec:MolQuenching}) and $n_e$ is the electron density of the considered plasma element at the time of reflection $t_r$.

The exact expression for the solid angle $\Delta \Omega_{\rm{sc}}$ can be written as
\begin{equation}
	\Delta \Omega_{\rm{sc}} = 2\pi \left( 1-\frac{r_{\rm{sc}}}{\sqrt{r_{\rm{sc}}^2+A_R \sin \alpha/\pi}} \right) \rm{,}  \label{fullomegasc}
\end{equation}
where the factor $\sin \alpha=h/r_{\rm{sc}}$ ($r_{\rm{sc}}=|{\bf r}_{\rm{sc}}|$) (see Figure \ref{fig:1}) is to account for the change in the effective area $A_R$ of the receiver antenna due to the angle $\pi/2-\alpha$ between the direction of the reflected wave and the receiver pointing vertically upward. For large $r_{\rm{sc}}$ we get the approximation
\begin{equation}
	\sqrt{	\Delta \Omega_{\rm{sc}}}  \approx \sqrt{A_R \sin \alpha} / r_{\rm{sc}} \rm{,}  \label{omegasc}
\end{equation}
so that $ {\rm d} U_{\rm{rcv}} \sim r^{-1} r_{\rm{sc}}^{-1}$. In practice, we use equation (\ref{omegasc}) and switch to the full expression (\ref{fullomegasc}) only for small $r_{\rm{sc}}$. This is to avoid unphysical increase in the ${\rm d} U_{\rm{rcv}}$ value when the scattering plasma is very close to the receiver. We analyze only the signal which arrives to the receiver antenna. Therefore,
we do not consider any characteristics of the antenna, i.e. its gain or directional dependence.

The propagation time of the radio wave from the point of reflection to the detector is
\begin{eqnarray}
 t -t_r	= \frac{1}{\sin \alpha} \int^{h}_{0} \frac{n(h')}{c } dh' \rm{.}
 \end{eqnarray}
From the above expression we derive
\begin{eqnarray}
 t_r	= t-\frac{r_{\rm{sc}}}{ c} n_h      \rm{,} \label{tr}
 \end{eqnarray}
where 
\begin{equation}
 n_h=\frac{1}{h} \int^{h}_{0} n(h') dh' \label{nh}
\end{equation} 
is a slowly varying function of $h$ and $c$ is the speed of light. The altitude $h$ of the plasma volume element ${\rm d}V$ can be derived from
\begin{equation}
	h=s \sin \theta_s -r_L \cos \varphi \cos \theta_s \label{hh} \rm{.}
\end{equation}

Assuming that the static plasma decays exponentially with the characteristic time $\tau$, the electron density $n_e$ is given by 
\begin{equation}
	n_e  = n_e^0 (s,r_L) \exp \left[-\frac{(s-s^{'})}{\tau c_s}\right] 	\Theta \left(s-s^{'}\right) 
	\rm{,} \label{ne-density}
\end{equation}
where 
\begin{equation}
	s^{'}=-c_s t_r= \frac{c_s}{ c} r_{sc} n_h- c_s t \rm{,}\label{sprim}
\end{equation}
is the position of the shower front at time $t_r$, i.e. at the moment of radio wave reflection from the plasma volume element ${\rm d}V$. To derive $s^{'}$, we have used the definition of the time zero as the moment in which the shower hits the ground. We use this definition throughout the paper, unless stated otherwise. $n_e^0 (s,r_L)$ denotes the electron density in the static plasma at the shower front at position $s$ and at distance $r_L$ from the shower axis, and $c_s$ is the speed of the shower front, respectively. In the following we assume that the shower front moves with the speed of light, i.e. $c_s=c$. The exponent is responsible for the plasma decay, whereas the Heaviside step function $\Theta(s-s^{'})$ assures that there is no plasma ahead of the shower front. 

Since the plasma lifetime $\tau$ at altitudes lower than 5 km is very short, i.e. $\tau \sim 15-40$ ns, it is enough to consider radar reflections only from the part of the ionization trail which is up to several $\tau$ behind the shower front to obtain numerically accurate results. In other words, the thickness of the plasma disk produced by the extensive air shower can be constrained to $n_{\tau}c_s \tau$.  
 In practice, we calculate $n_{\tau}$ for each altitude separately, so that less than $5\%$ of the total plasma electrons is not being accounted for. 
 
The signal received by the antenna at a given time is a sum of the signals scattered at different times, from different parts of the plasma disk, and from different altitudes. These individual contributions interfere with each other and only an integral over the whole volume $V(t)$ from which they arrive simultaneously gives us the correct value. This volume is, in general, time-dependent and it can extend over a wide range of altitudes (even several kilometers due to the time compression of the received signal).
 Combining equations (\ref{Uinc}) and (\ref{dUrcv}), and integrating ${\rm d}U_{\rm{rcv}} / {\rm d} V$ over $V(t)$, we get the following expression for the total electric field strength of the radio wave at the receiver at time $t$
\begin{equation}
	U_{\rm{rcv}}(t) =  \iiint\limits_{V(t)} \frac{\mathrm{d} U_{\rm{rcv}} (t,s,r_L,\varphi)}{\mathrm{d} V} \mathrm{d}V  \rm{,} \label{total}
\end{equation}
where
\begin{eqnarray}
\frac{{\rm d}U_{\rm{rcv}} (t, s, r_L, \varphi )}{{\rm d}V} &=&  U_{T} \sqrt{ \left( \frac{\omega}{\nu_c} \right)^2  \frac{3\sigma_T( 1+\cos^2 \theta )   }{16 \pi  }  } \frac{\sqrt{G_T \Delta \Omega_{sc}}}{|{\bf r}|}  n_e    \nonumber \\
& \times & e^{i(\omega t + \phi_0)}  e^{-i\int_{{\bf r}} n \hspace{0.1cm} {\bf k} \cdot {\rm d}{\bf r}} e^{ - i\int_{{\bf r_{sc}}} n \hspace{0.1cm} {\bf k_{sc}} \cdot {\rm d} {\bf r_{sc}} }  \label{Ucontr}
\end{eqnarray}
is the contribution to the radar echo from the plasma volume element ${\rm d}V = r_L {\rm d}r_L {\rm d}\varphi {\rm d}s$  with coordinates ($s$, $r_L$, $\varphi$). The details of $U_{\rm{rcv}}(t)$ calculation can be found in the Appendix.

Equation (\ref{Ucontr}) displays the expected dependence of the field strength on the distances $r=|{\bf r}|$ and $r_{\rm{sc}}=|{\bf r_{sc}}|$, and the spatial phase which is due to the transmitter-plasma-receiver optical path of the radio wave. For a high altitude we have roughly $r \approx r_{\rm{sc}} $ and $\sqrt{\Delta \Omega_{\rm{sc}}} \sim 1/r$, so the field strength diminishes like $r^{-2}$. This is equivalent
to an $r^{-4}$ dependence in the received power.

\subsection{Effective cross section}

Alternatively, the radar reflection can be described in terms of the effective cross-section, which is a measure of the target equivalent area of an ideal scattering surface. The effective shower cross-section can be defined, by analogy with \cite{bibe:gorham}, in the following way
\begin{equation}
	\sigma_{\rm{eff}} (t)   =  \left|  \int_{V(t)}  e^{-i\int_{{\bf r}} n \hspace{0.1cm} {\bf k} \cdot \mathrm{d}{\bf r}} e^{ - i\int_{{\bf r_{sc}}} n \hspace{0.1cm} {\bf k_{sc}} \cdot \mathrm{d} {\bf r_{sc}} } n_e  \sqrt{  \left(\frac{\omega}{\nu_c} \right)^2  \frac{\mathrm{d} \sigma_T}{ \mathrm{d} \Omega}} \mathrm{d} V  \right|^2
 \rm{.} \label{eff-cross}
\end{equation}
Note that after removing the geometrical factor $\sqrt{\Delta \Omega_{\rm{sc}}} /|{\bf r}|$ of the considered radar system  from the definition of ${\rm d}	U_{\rm{rcv}} (t,s,r_L,\varphi)/ {\rm d}V$ in equation (\ref{Ucontr}), we obtain $\sigma_{\rm{eff}} (t) \sim |U_{\rm{rcv}}(t)|^2$.

 $\sigma_{\rm{eff}}(t)$, which is defined by equation (\ref{eff-cross}), has the meaning of a cross-section only when the radio transmitter and receiver are sufficiently far away from the scattering plasma or the volume of the scattering plasma is very small itself. These conditions will not always be met. In reality, the volume from which the scattered radio waves arrive simultaneously to the detector can have a considerable size for small viewing angles between the shower axis and the direction of $\bf r_{sc}$. This is caused by the time compression of the reflected signal. The estimation of radar cross-section of a shower given by $\sigma_{\rm{eff}}(t)$ is used in the following only for analysis of signal compression separately from the geometrical factor.

\subsection{Maximum of the received signal} 
 
\label{SecMax}

The ratio of the instantaneous power $P_R(t) = \operatorname{Re}( U_{\rm{rcv}}(t))^2/Z_0$ received by the detector antenna to the power emitted by the transmitter $P_T = 4 \pi U_T^2 /G_TZ_0$ is equal to 
\begin{equation}
		P_R(t)/P_T = R^2(t) \rm{,} \label{PR}
\end{equation}
where $Z_0 \approx 120 \pi$ $\Omega$ is the impedance of free space and $R(t)$ is defined by
\begin{eqnarray}
	R(t) &=& \operatorname{Re}(U(t)) \rm{,} \\
	U(t) &=& \sqrt{\frac{G_T}{4 \pi}}\frac{U_{\rm{rcv}}(t)}{U_T} \rm{.}  \label{Ut}
\end{eqnarray}

The instantaneous signal strength $U(t)$  is a useful dimensionless quantity which facilitates deriving the maximum of the received power. Its real part defines the waveform $R(t)$, which is proportional to the electric field strength received by the detector and is used in the Fourier analysis to obtain the power spectrum of the recorded signal. Note that, for large distances of the scattering centers to the detector, $P_R/P_T$ is proportional to the effective area of the receiver antenna and to the transmitter gain, i.e. $P_R/P_T \sim G_T A_R$. 

The shape of the waveform $R(t)$ and thus of the received power $P_R(t)$ depend, via the electric field $U_{\rm{rcv}}(t)$, on the initial phase $\phi_0$ of the transmitter. Hence, a different choice of the $\phi_0$ value will change both the moment at which the detector observes the maximum power and the maximum value of this power $P_{R,m}$. For each shower, we scan through all possible $\phi_0$ values to maximize $P_{R,m}$. Let us denote this maximum $P_{R,m}$ value by $P_{R,\rm{max}}$ and the corresponding time-dependent function $P_R(t)$ by $P_{R,\rm{max}}(t)$. From the equations (\ref{PR})-(\ref{Ut}), we can deduce that $P_{R,\rm{max}}/P_T$ is equal to the square of the maximum value of the function $|U(t)|$. The maximum received power for a given shower is represented by $P_{R,\rm{max}}$ and its time dependence by $P_{R,\rm{max}}(t)$.

\subsection{Frequency upshift}

 \begin{figure*}[t]
  \centering
  \includegraphics[width=1\textwidth]{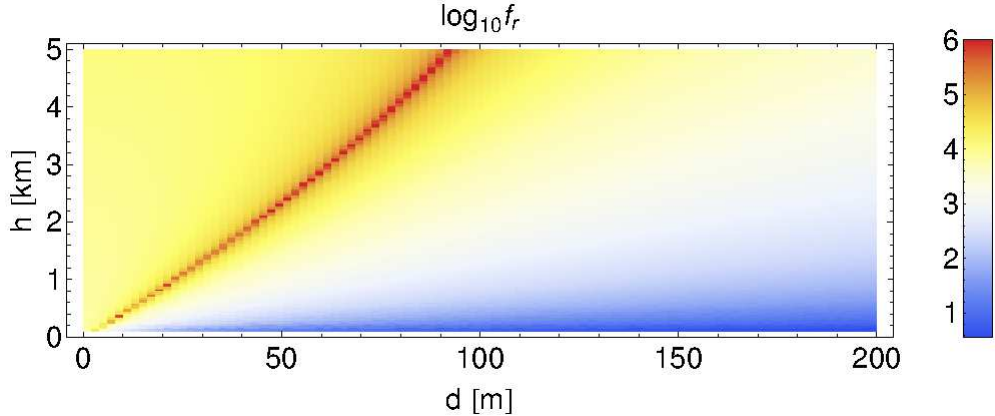}
  \caption{Distribution of the factors by which the emitted frequency is up-shifted ($f_r$) for the radio wave scattered off different parts of the shower disk. The altitude of the disk element is given by $h$, whereas its distance to the receiver in the horizontal plane is $d$. A vertical shower heading towards the transmitter is considered.}
  \label{fig:upshift}
 \end{figure*}  

Despite the fact that the radio wave is scattered on a non-moving plasma, the shower ionization front, and thus the wave scattering region, moves with a relativistic velocity. Therefore, a Doppler effect should be observed in the received signal. Figure \ref{fig:upshift} shows the factors by which the emitted frequency is up-shifted ($f_r$) for the radio wave scattered on different parts of the disk-like plasma produced by a vertical shower heading towards the transmitter. The altitude of the plasma element is given by $h$, whereas its distance to the receiver in the horizontal plane is $d$. 

The frequency upshift depends on the wave direction and on the refractive index of the air. It is largest for the case in which the viewing angle coincides with the Cherenkov angle. The factor $f_r$ can be high enough to upshift a MHz signal into the GHz range. Therefore, it might be possible to observe the radar echo in the GHz range using a microwave detector \cite{bibe:MIDAS,bibe:AMBER,bibe:EASIER,bibe:CROME,bibe:smida} supplemented with a high-power MHz transmitter. However, our results (see Section \ref{ressim}) show that the typical value of $f_r$ is much lower. The frequency upshift derivation can be found in \cite{bauer}. 

\section{Air shower plasma}

\label{air-plasma}

\subsection{Plasma density}

 \begin{figure}[bt!]
  \centering
  \includegraphics[width=1\textwidth]{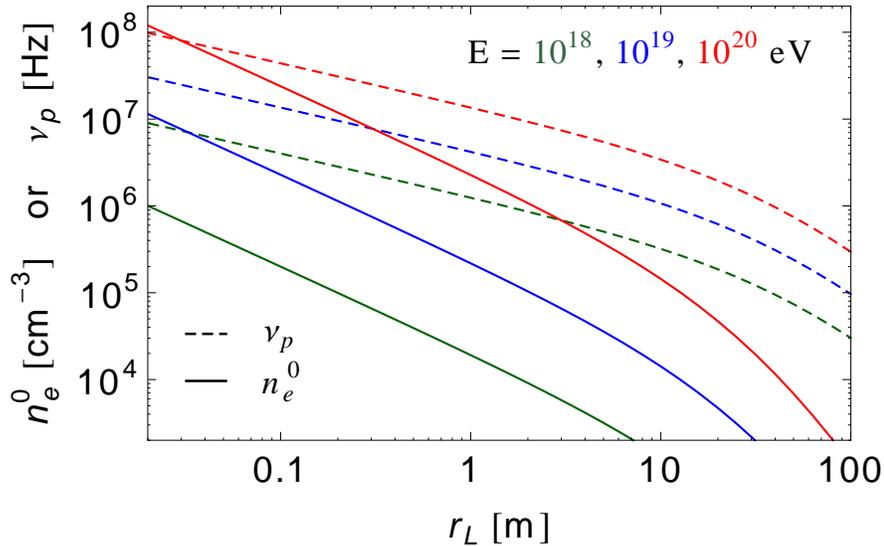}
  \caption{Radial dependence of the initial electron density $n^0_e$ produced by a shower at its maximum for vertical showers with energies $10^{18}$, $10^{19}$, and $10^{20}$ eV (solid curves from the bottom to the top, respectively). Also shown is the radial dependence of the plasma frequency $\nu_p = \omega_p /2\pi$ for each energy (dashed curves).}
  \label{fig:plasma}
 \end{figure}

The initial electron density of the plasma $n^0_e$, i.e. the plasma density immediately after passing of the shower front, produced by the high-energy shower particles in the air, is estimated using the average longitudinal profile of proton showers parametrized by the Gaisser-Hillas function \cite{Gaisser-Hillas} and assuming the G\'ora function \cite{bibe:gora} as the lateral distribution. The G\'ora function corresponds to the lateral distribution of the shower energy deposition in the air. We assume that each shower particle deposits on average 2.3 MeV per traversed g/cm$^{2}$ and that all of the deposited energy goes into ionization. The mean energy per ion pair production is 33.8 eV \cite{Rossi}. For the distances from the shower axis ($r_L$) smaller than 1 cm, the electron density is assumed to be constant and equal to the value of the density at $r_L=1$ cm  to avoid its unphysical increase. This assumption does not affect our results, since the number of electrons within this region is below $0.1\%$ of their total number.

While the maximum particle count of the shower is described by the Gaisser-Hillas function, the plasma density depends also on the local atmospheric density. Therefore, the point of the shower maximum does not necessarily correspond to the point of the highest plasma density. In fact, there is an offset in the altitude between these two points. For vertical showers with energy higher than $10^{18}$ eV, the maximum of the plasma density is about 400-500 m below the shower maximum. The difference between the  plasma density at the shower maximum and at the plasma maximum is less than a few percent. The plasma densities as well as the altitude offsets between the shower and the plasma maximum which we obtain, are in a good agreement with the results of more detailed calculations of the plasma production \cite{bibe:n1,bibe:n2}. 

The plasma frequency is defined by
\begin{equation}
	\omega_p = \sqrt{\frac{e^2n_e}{\epsilon_0 m_e}} \approx 5.64 \times 10^4 \sqrt{n_e} \hspace{0.2cm} \rm{[Hz] ,}
\end{equation}
where $e$, $m_e$, and $\epsilon_0$ are the elementary charge, the electron mass and the permittivity of the free space, respectively. The electron density $n_e$ is given in cm$^{-3}$. The plasma frequency corresponds to the characteristic electrostatic oscillation frequency of an electron in the plasma in response to a small charge displacement. The value of $\omega_p$ is of high importance in determining scattering properties of the plasma.

Examples of the lateral distribution of the initial electron density $n^0_e$ (solid curves) and the corresponding plasma frequency $\nu_p = \omega_p /2\pi$ (dashed curves)  are shown in Fig. \ref{fig:plasma}. The three curves presented correspond to the plasma, at the shower maximum, produced by the vertical showers with energies $10^{18}$, $10^{19}$, and $10^{20}$ eV.

The electron density is highest at the shower axis, reaching up to several times $10^8$ cm$^{-3}$ (for showers with energy $10^{20}$ eV), and decreases steeply with the distance to the shower axis $r_L$. For $r_L=0.1$ m, the electron density is about one tenth of the density at the shower axis, whereas for $r_L=$1 m it falls down to one-hundredth of the axis value. The radial dependence of the electron density agrees well with the dependence derived from CORSIKA simulations \cite{borodai}. 

\subsection{Plasma lifetime}

 \begin{figure}
  \centering
  \includegraphics[width=1\textwidth]{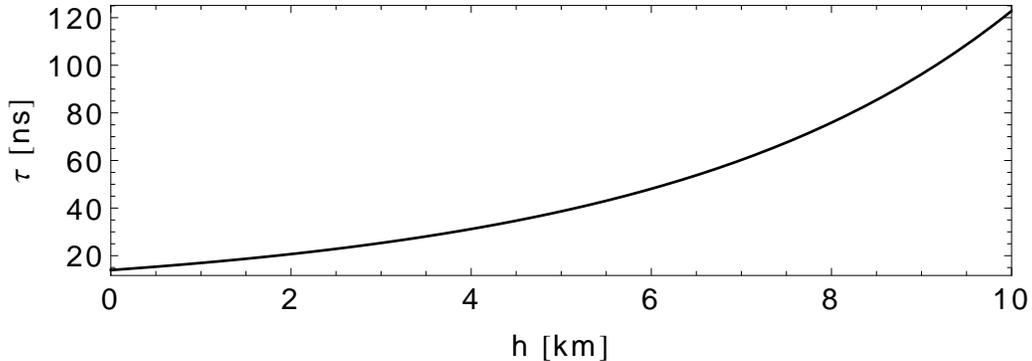}
  \caption{Plasma lifetime in the air as a function of altitude.}
  \label{fig:tau}
 \end{figure}

\label{plasma-lifetime}
 
Due to the low number of electrons in the plasma produced by a shower in the air relative to the air molecule density, electron attachment  (as opposed to recombination) is the principal mechanism for electron removal from the plasma. Electron attachment occurs by formation of an unstable state of the negative ion $\rm{O}^-_2$, which subsequently may re-emit an electron (auto-detachment) \cite{chanin}. However, a collision with a third body can remove some energy, causing transitions to non-autoionizing levels of the $\rm{O}^-_2$ ion.
For $n_e < 10^{12}$ cm$^{-3}$, as is always the case for the plasma produced by showers (see Figure \ref{fig:plasma}), the deionization process is dominated by the
three-body attachment of electron to oxygen molecules \cite{bibe:vidmar}:
\begin{eqnarray}
	e^- + \rm{O}_2 + \rm{O}_2 &\rightarrow& \rm{O}^-_2 + \rm{O}_2 \rm{,} \\ 
		e^- + \rm{O}_2 + \rm{N}_2 &\rightarrow& \rm{O}^-_2 + \rm{N}_2 \rm{.}	
\end{eqnarray}
 
The deionization process described above leads to an exponential decay of the plasma. The characteristic decay time $\tau$, i.e. the time required for a plasma to decrease in concentration by a factor of $1/e$, is given by \cite{bibe:nijdam}
\begin{equation}
	\tau = N_{\rm{m}}^{-2} \left( k_{\rm{att}1}[\text{O}_2]^2  + k_{\rm{att}2} [\rm{O}_2] [\rm{N}_2] \right)^{-1} \rm{,}
\end{equation}
where $[\text{O}_2]=0.209476$ and $[\text{N}_2]=0.78084$ are fractions of oxygen and nitrogen in the atmosphere, while $k_{\text{att}1}=2\times10^{-30}$ cm$^6$ s$^{-1}$ and $k_{\text{att}2}=8\times 10^{-32}$ cm$^6$ s$^{-1}$ \cite{bibe:nijdam}. $N_{\text{m}}$ is the total number density of atmospheric molecules (nitrogen, oxygen, carbon-dioxide and trace gases). The altitude-dependent molecular number density $N_{\text{m}}$ is derived from the ideal gas law assuming the U.S. Standard Atmosphere model \cite{bibe:USstandard} values of the air temperature and pressure.

The dependence of the plasma lifetime on altitude is shown in Fig. \ref{fig:tau}. The plasma lifetime changes from about 15 ns at the sea level to 40 ns  at the altitude of 5 km and 120 ns at 10 km.

\subsection{Effective collision frequency} 

The electron collision frequency is an important quantity determining absorption of radio waves and quenching their reflection off the atmospheric plasma. It is defined as the number of collisions  of an electron with heavy particles of the plasma and other electrons occurring in 1 s.

For a tenuous plasma (with $n_e<10^8$/cm$^3$, see Fig. \ref{fig:plasma}), which we consider, the density of free electrons is much lower than the density of neutral molecules. Therefore, the collision frequency of an electron with neutral molecules is much larger than the electron-electron and electron-ion collision frequencies. Consequently, we neglect collisions with electrons and ions and consider only electron collisions with neutral molecules, i.e. with $\text{N}_2$ and $\text{O}_2$.

The velocity-dependent collision frequency of electrons is defined as \cite{shkarofsky}
\begin{equation}
	\nu_{\text{m}}(v)=N_{\text{m}} v Q_{\text{m}}(v) \rm{,} \label{num}
\end{equation}
where $N_{\text{m}}$ is the gas molecule number density, $v$ is the incident electron velocity and $Q_{\text{m}}(v)$ is the velocity dependent momentum transfer cross-section given by
\begin{equation}
	Q_{\text{m}}(v)=[\text{N}_2] Q_{\text{m}}^{\text{N}_2}(v) + [\text{O}_2] Q_{\text{m}}^{\text{O}_2}(v)  \rm{.}
\end{equation}
The values of the momentum transfer cross-sections for electron impact on $\text{N}_2$ and O$_2$, i.e. $Q_{\text{m}}^{\text{N}_2}(v)$ and $Q_{\text{m}}^{\text{O}_2}(v)$, can be found in \cite{itikawa06} and \cite{itikawa09}. For altitudes at which extensive air showers develop in most part, i.e. $h<10$ km, the mean value of $\nu_m$ is of the order of a few THz. Thus, the collision frequency is much larger than frequency of the radio waves considered in this paper. 

The collision frequency defined by equation (\ref{num}) can be used only for a monoenergetic beam of particles. To obtain the velocity-independent effective collisional frequency $\nu_c$, we have to appropriately average $\nu_{\text{m}}(v)$ over the velocity distribution of the plasma electrons. Assuming the Maxwellian distribution of $v$, the effective collision frequency is defined by the following equation
\cite{itikawa} 
\begin{equation}
	\nu_c^{-1} = \frac{8}{3\sqrt{\pi}N_{\rm{m}}} \left(\frac{m_e}{2k_BT_e} \right)^{5/2} 
	\int_0^{\infty} \frac{v^3}{Q_{\text{m}}(v)}\exp\left(-\frac{m_e v^2}{2k_BT_e}\right)dv \rm{,} \label{nueff}
\end{equation}
where $T_e$ is the temperature of electrons and $k_B$ is the Boltzmann constant. 

Strictly speaking, the distribution function of the plasma induced by a shower in the air is non-Maxwellian \cite{bibe:n1,bibe:n2}. 
However, for our purposes we approximate this non-Maxwellian distribution by the thermal one. The  effective temperature $T_e$ of the new distribution is defined in such a way that the mean kinetic energy per particle is the same as in the original (non-Maxwellian) distribution. We adopt $T_e=1.5 \times 10^5$ K from \cite{bibe:n1}.

The resulting effective collisional frequency $\nu_c$ is a monotonic function of altitude. It decreases from about 5 THz at the sea level to less than 2 THz at the altitude of 10 km.  

\label{sub:coll}

\subsection{Absorption and refractive effects of the plasma}

When a radio wave propagates in the plasma, it suffers attenuation because of electron collisions with the gas molecules. In the absorption process, energy of the electromagnetic wave is converted into thermal energy of the gas via coupling to the plasma electrons:
the plasma electrons are first accelerated by the electric field of the radio wave and then collide with gas molecules, thereby transferring energy from the electromagnetic wave to the gas. This process is irreversible, thus it prevents re-emission of radio waves and so scattering of the radar wave is quenched.

The absorption of the electromagnetic waves in plasma has a resonant character; the absorption coefficient is largest when the values of the incoming radio frequency $\omega$, plasma frequency $\omega_p$ and collision frequency $\nu_c$ are very similar \cite{bibe:Santoru,bibe:Zhang,bibe:Zhongcai}. For a given plasma density (fixed $\omega_p$), the absorption is maximized for $\omega \approx \nu_c$ and decreases for larger or smaller collision frequencies. When $\nu_c \gg \omega$, the plasma absorbs only a very small fraction of the electromagnetic wave power: the absorption process is inefficient because electrons acquire very little energy before they collide with molecules. When the opposite is true (i.e. $\nu_c \ll \omega$), the collision rate is so low that the electrons simply oscillate unimpeded in the electric field of the radio wave and little wave energy is absorbed. 

For $\nu_c \gg \omega$, as in our case, the electrical conductivity of the plasma takes the form of the conductivity for a static field \cite{itikawa}, i.e.
\begin{equation}
	\sigma=\frac{\epsilon_0 \omega_p^2 }{\nu_c} \rm{.} \label{cond}
\end{equation}
Using the above expression, we can derive the refractive index of the plasma, which is
\begin{equation}
	n_p=\sqrt{1 - \frac{i \sigma }{\omega \epsilon_0}} \approx 1 -\frac{\omega_p^4}{8\nu_c^2\omega^2} - i \frac{\omega_p^2}{2\nu_c\omega} \rm{.} \label{eq:n}
\end{equation}
Accordingly, the propagation dependence of the electromagnetic field is
\begin{equation}
 e^{i(\omega t - nk_0r)} =e^{k_0n_ir}e^{i(\omega t-k_0n_rr)} \rm{,} \label{propag}
\end{equation}
where $k_0=\omega / c$ is the free-space wave number, whereas $n_r$ and $n_i$ are the real and the imaginary part of $n$, respectively. 
The absorption coefficient $\alpha_{\text{abs}}$ follows from (\ref{propag}), i.e.
\begin{equation}
	\alpha_{\text{abs}}=-2k_0n_{\text{i}}=\frac{\omega_p^2}{\nu_c c} \rm{.}
\end{equation}

The highest value of $\alpha_{\text{abs}}$ is attained for the highest plasma density and the lowest collision frequency. An upper limit on the absorption coefficient $\alpha_{\text{abs}} < 3$ dB/km can be obtained by adopting $\nu_p =\omega_p /2\pi \approx 100$ MHz and $\nu_c \approx 2$ THz. The first value corresponds to the frequency of the plasma, at the shower maximum, produced by the $10^{20}$ eV shower close to its axis (see Fig. \ref{fig:plasma}). The second is the collision frequency at an altitude of 10 km. 

The region with the highest plasma density has a size of only several centimeters in diameter and several meters in length. Outside this region, the electron density drops by at least one order of magnitude and thus the absorption coefficient becomes lower than $0.3$ dB/km. It follows that the absorption of the radio wave in the plasma produced by the shower in the air can be neglected. This might not be the case only for showers heading towards the receiver. Due to the small difference between the speed of light in the air and the shower front velocity, the reflected radio wave can stay within the densest part of the plasma through all of the propagation time to the receiver. Nevertheless, even in the worst case, i.e. when the radio wave propagates over a distance of 5 km (the altitude limit of the signal integration, see Section \ref{Assump}) through the densest plasma, the decrease in the radio wave energy will be lower than 15 dB.
It will be about 15 dB for a 10$^{20}$ eV shower but only 1 dB for a $10^{19}$ eV shower.

 Similarly, we can obtain the phase shift of the radio wave after propagating a distance $l$ through the plasma, i.e.
\begin{equation}
	\Phi_{ps}=-k_0(n_r-1)l= \frac{\omega_p^4}{8\nu_c^2\omega c}l \rm{.}
\end{equation}
The phase shift of the radio wave per unit distance of propagation through the plasma $\Phi_{ps}/l$ decreases with the radio wave frequency. Since we do not consider radio wave frequencies lower than $\nu=1$ MHz, an upper limit of only 0.1 deg/km can be derived. The phase shift effect can be neglected.

\label{sub:absorption}
 
\subsection{Scattering off the plasma}  
\label{scatter}

For a collisionless plasma ($\nu_c=0$), the total reflection from the non-moving interface between the air and the homogeneous plasma occurs when $\omega \leq \omega_p$, and the reflected power drops for $\omega > \omega_p$. If collisions are present, then the radio wave is not totally reflected, even for $\omega < \omega_p$, and the wave can penetrate the plasma volume. The reflected power is greatly reduced as the electron collision frequency $\nu_c$ increases (see Figure 4 in \cite{bibe:Santoru}). For the specular reflection to occur, it is necessary not only that the radio wave frequency $\omega$ is lower than the plasma frequency $\omega_p$ ($\omega < \omega_p$) but also that it is much higher than the collision frequency $\nu_c$ ($\omega \gg \nu_c$). In our case, these conditions are not fulfilled. Even though the first condition might be met by the plasma produced by the shower at its axis (for a low radar frequency, i.e. $\nu<16$ MHz), the second one is never fulfilled.

The refractive index of the plasma for arbitrary values of $\omega$, $\omega_p$ and $\nu_c$ is \cite{bibe:Santoru}
\begin{equation}
	n_p = \left(1-\frac{\omega_p^2}{\omega^2+\nu_c^2}-i\frac{\nu_c}{\omega}\frac{\omega_p^2}{\omega^2+\nu_c^2}   \right)^{1/2} \rm{,}
\end{equation}
whereas the fraction of the incident power that is reflected from the non-moving interface of the homogeneous plasma is given by the Fresnel formula
\begin{equation}
	R_{p} = \left| \frac{n-n_p}{n+n_p} \right|^2 \rm{.}
\end{equation}

Since the plasma induced by meteors at altitudes $80-120$ km reaches the thermal equilibrium in a time much shorter than the plasma lifetime, we can assume that the electron temperature is $T_e\approx200$ K and obtain an upper limit for the collision frequency $\nu_c$ of only 1 MHz. Moreover, assuming that the diameter of the meteor trail is about 20 m and using radar frequency of $\nu=10$ MHz we can derive the reflection coefficients, which are $R_p\approx4\times 10^{-5}$ for the ionization line density of $10^{11}$ cm$^{-1}$ (which represents an underdense case) and $R_p\approx0.97$ for the line density of $10^{13}$ cm$^{-1}$ (overdense case). These numbers agree very well with our expectation based on reflective properties of underdense and overdense regions.

In the case of a plasma produced by a shower, we expect the strongest reflection to occur for $\nu_c=2$ THz, $\nu=1$ MHz and $\nu_p=100$ MHz for the range of parameters considered here.  For this set of parameters we get $R_p\approx6\times10^{-5}$. The plasma frequency $\nu_p=100$ MHz corresponds to the plasma produced by a $10^{20}$ eV shower close to its axis (the densest plasma region). The reflection coefficient drops to $R_p \approx 10^{-11}$ at a distance of $r_L=1$ m from the shower axis. As expected, the shower plasma has to be always treated as underdense. This is caused by the high density of air at altitudes at which the shower plasma is created, i.e. altitudes lower than 10 km, and the resulting high value of $\nu_c$.

Let us now consider reflection from the plasma disk left behind the shower front at small angle to the shower axis. In such a case we have to analyze the radio wave reflection from the moving ionization front (plasma boundary) under condition that the electrons of the plasma are static. For simplicity, let us restrict to the collisionless plasma. Then, the condition for the total reflection $\omega < \omega_p$ has to be replaced by $\omega' < \omega_p$ \cite{bibe:stephanov,bakunov3,samarai}, where
\begin{equation}
	\omega' = \omega \frac{n \beta +\cos \theta_0 }{\sqrt{1-n^2\beta^2}}
\end{equation}
is the frequency of the incoming radio wave in the coordinate system in which the plasma boundary is stationary. The parameter $\beta=c_s/c$ is the ratio of the ionization front velocity to the speed of light in vacuum and $\theta_0$ is an angle of incidence of the radio wave on the plasma boundary. The frequency $\omega'$ increases to infinity when $n\beta \rightarrow 1$ and the plasma becomes transparent to arbitrarily low-frequency incident radiation, i.e. it becomes underdense for all frequencies of the incoming wave. This situation holds for $n\beta \geq 1$, when a reflected wave in the shower forward direction cannot exist because it would be immediately caught by the shower front. In its place a second transmitted wave with a highly reduced strength is formed \cite{bibe:stephanov}. This so-called transmitted-back-scattered wave follows the ionization front while staying behind it. The frequency of this wave, similarly to the reflected one (when it exists), is highly upshifted \cite{bibe:stephanov}.

The shower front moves with the speed of the highest-energy particles in a shower and thus exceeds the local speed of light ($n\beta >1$) at all altitudes of relevance. Therefore, in the case of scattering of the radio waves incoming at small angles to the shower axis, one can not treat the plasma as overdense. Moreover, unlike the case of reflection from the side of the ionization trail, where the frequency does not change, the frequency of the transmitted-back-scattered radio wave will be upshifted. One can then expect an enhancement of the signal scattered backwards due to its time compression. 

It follows that we can not use an analogy with the reflective behaviour of the overdense ionization trails produced by meteors high in the atmosphere and apply it to the plasma produced by the shower at much lower altitudes. For the overdense meteor trails, secondary scattering of the radio wave on electrons is important, the refractive index of the plasma becomes imaginary and the total reflection of the incoming radio wave occurs. An overdense meteor trail can be modeled as a thin metallic cylinder and the corresponding radar cross-section can be used. In contrast, the plasma induced by a shower has to be treated as underdense in all cases. The radio wave can penetrate the ionized region and the reflections of the radio wave are caused by its scattering on individual electrons. The total scattered signal is a sum of these individual contributions. The above conclusion does not change even if we consider scattering on a plasma-density gradient (as is the case for air shower plasma) rather than on a sharp boundary (steep plasma-density gradient).  

Therefore, treating the plasma as underdense and using the Thomson cross-section, with a correction for molecular quenching (see Section \ref{sec:MolQuenching}), is justified for radar scattering in case of air showers.

\subsection{Molecular quenching}

\label{sec:MolQuenching} 

Following \cite{bibe:filonenko} let us write the equation of motion for an electron in the electromagnetic field of the radio wave, i.e.
\begin{equation}
	\ddot{z} + \nu_c \dot{z} = \frac{eE_0}{m_e} \cos (\omega t -k x) \rm{,}   
\end{equation}
where $E_0$ is the amplitude of the electric field of the radio wave and the term $\nu_c \dot{z}$ represents the drag force arising in collisions with neutral molecules. The steady-state solution is proportional to the driving force and for $\nu_c \gg \omega$ we get 
\begin{equation}
	z(t)  = \frac{eE_0}{m_e \omega \sqrt{ \omega^2+ \nu_c^2 }} \sin (\omega t -k x) \approx \frac{eE_0}{m_e \omega \nu_c} \sin (\omega t -k x) \rm{.}  \label{eq-osc}
\end{equation}

The Thomson cross-section $\sigma_T$ describes scattering of the electromagnetic wave on a non-relativistic, free electron in vacuum, i.e. when there are no collisions with other particles. Thus we have $\nu_c=0$, the drag force term disappears, and the amplitude in equation (\ref{eq-osc}) becomes equal to $eE_0 / m_e \omega^2 $.

The cross-section $\sigma_{\text{sc}}$ for scattering of the electromagnetic wave on a free electron, by definition, is proportional to the average power re-radiated by the electron. This power, according to the Larmor formula, is proportional to the average value of the square of electron acceleration, i.e. $<\ddot{z}(t)^2 >$. It follows that $\sigma_{\text{sc}} \sim <\ddot{z}(t)^2 >= (eE_0 \omega / m_e \nu_c )^2$ and similarly $\sigma_T \sim (eE_0 / m_e  )^2$. Therefore, we have
\begin{equation}
	\sigma_{sc} = \left(\frac{\omega}{\nu_c} \right)^2 \sigma_{T} \rm{.}
\end{equation}

The net effect of the electron collisions with neutral molecules is a reduction of the electron acceleration in the radio wave field and thus a decrease in the power re-radiated by the electron. This effect is called the molecular quenching. It reduces the power recorded in the receiver antenna by 8-12 orders of magnitude for the radio frequency $\nu = \omega / 2\pi$ in the range of 1-100 MHz. The decrease is larger at lower frequencies. 

To take into account the above effect, we scale the received signal by the factor $\omega / \nu_c$ (see equations (\ref{dUrcv}), (\ref{Ucontr}) and (\ref{Ucontr2})). In other words, we use the Thomson cross-section with a correction for molecular quenching to describe the scattering properties of the plasma electrons.

\section{Simulations and analysis} 

\label{SimAnalyze}

\subsection{Assumptions}

\label{Assump}

An extensive set of simulations of radar detection of air showers were made, as outlined in Sections \ref{modeling} and \ref{air-plasma}.
The simulations were performed assuming that the effective area of the receiver antenna is $A_R$=1 m$^2$ and the ground-based transmitter ($h_T=0$) emits signal isotropically into the whole upper hemisphere (i.e. $G_T=2$). We also assume that the receiver points vertically upward and the solid angles ($\Delta \Omega_{\text{sc}}$) at which it is seen from each point of scattering are taken into account. Thus, we are not constrained to the far field approximation. Both the transmitter and the receiver are located at the sea level. We analyze only the signal which arrives to the receiver antenna. We do not consider any characteristics of the antenna, i.e. the receiver is assumed to be ideal and its efficiency is independent of the frequency of the radar echo.

     \begin{figure}[t]
  \centering
    \includegraphics[width=0.9\textwidth]{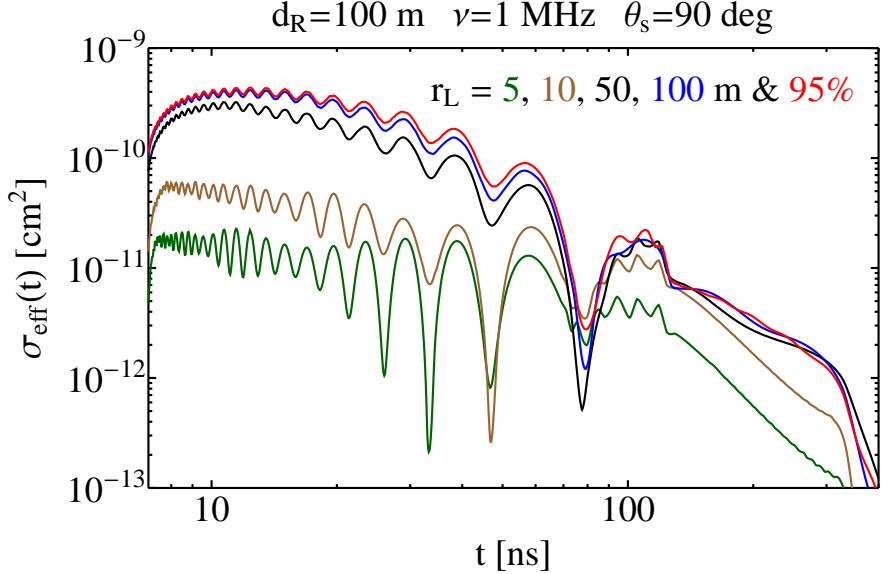}
  \caption{Dependence of the radar effective cross-section obtained after integration over the plasma volume up to the different distances $r_L$ from the shower axis as a function of observation time, calculated for a vertical shower with energy $10^{18}$ eV heading towards the transmitter. The distance between the shower core and the receiver is $d_R=100$ m and the frequency of the incident radar wave is $\nu=1$ MHz. 
   The curves from the bottom to the top represent radar cross-sections obtained by integration over the plasma volume up to the distances $r_L$=5, 10, 50, 100 m from the shower axis, respectively. The uppermost, red curve is the radar cross-section calculated taking into account 95\% of the total plasma electrons.   
   }
  \label{fig:sigma-rL}
 \end{figure}
 
Since the received power of the radar echo is strongly diminished by the geometrical factor of $r^{-4}$, the strongest signal is usually obtained from altitudes close to the ground level. Therefore, we limit integration of the signal contributions up to the altitude of 5 km and use the collision frequency at 1 km, i.e. $\nu_c=4.5$ THz, to correct the Thomson cross-section for molecular quenching. Since the collision frequency varies from about 5 THz to 3 THz in the altitude range of 0 to 5 km, the maximum received power can be higher than the calculated one by a factor of 2.25 (+3.5 dB). This is true for the power maximum ($P_{R,\rm{max}}$) originating at high altitudes, i.e. for inclined showers with the shower core located far away from the receiver and transmitter antennas (see Section \ref{ressim}). We do not simulate the whole radar echo but rather end the simulation at the moment at which the signal decreases considerably due to the decay of the plasma. From now on, unless stated otherwise, the time $t=0$ coincides with the moment at which the shower hits the ground.

Figure \ref{fig:sigma-rL} shows the dependence of the radar effective cross-section obtained after integration over the plasma volume up to the different distances $r_L$ from the shower axis as a function of observation time, calculated for a vertical shower with energy $10^{18}$ eV heading towards the transmitter. The shower core-receiver distance is $d_R=100$ m and the frequency of the incident radar wave is $\nu=1$ MHz. The logarithmic time-scale was chosen to show an early part of radar cross-section in more detail. The curves from the bottom to the top represent radar cross-sections obtained with integration over the plasma volume up to the distances $r_L$=5, 10, 50, 100 m from the shower axis, respectively. The uppermost, red curve is the radar cross-section calculated taking into account 95\% of the total plasma electrons.
   
As can be seen, the innermost (densest) part of the ionization trail does not provide the dominant contribution to the reflected power. In fact, for $\nu=1$ MHz ($\lambda\approx 300$ m), contributions to the signal sum up coherently and most of the signal is produced up to the distance of $r_L \approx 100$ m from the shower axis. This region contains about $80\%$ of the total plasma particles. For higher radio frequencies the coherence length is smaller, thus it is of high importance to take into account the destructive interference and to sum individual contributions generated at larger distances from the shower axis. Therefore, we sum up the scattering contributions up to the lateral distance which contains $95\%$ of the total plasma electrons, i.e. up to several hundred meters from the shower axis. This distance is smaller for a vertical shower and increases with the shower inclination angle.  

\subsection{Signal analysis}
\label{ressim}

\begin{figure}[btps!]
  \vspace{-0.4cm}
  \centering
 \includegraphics[width=0.75\textwidth]{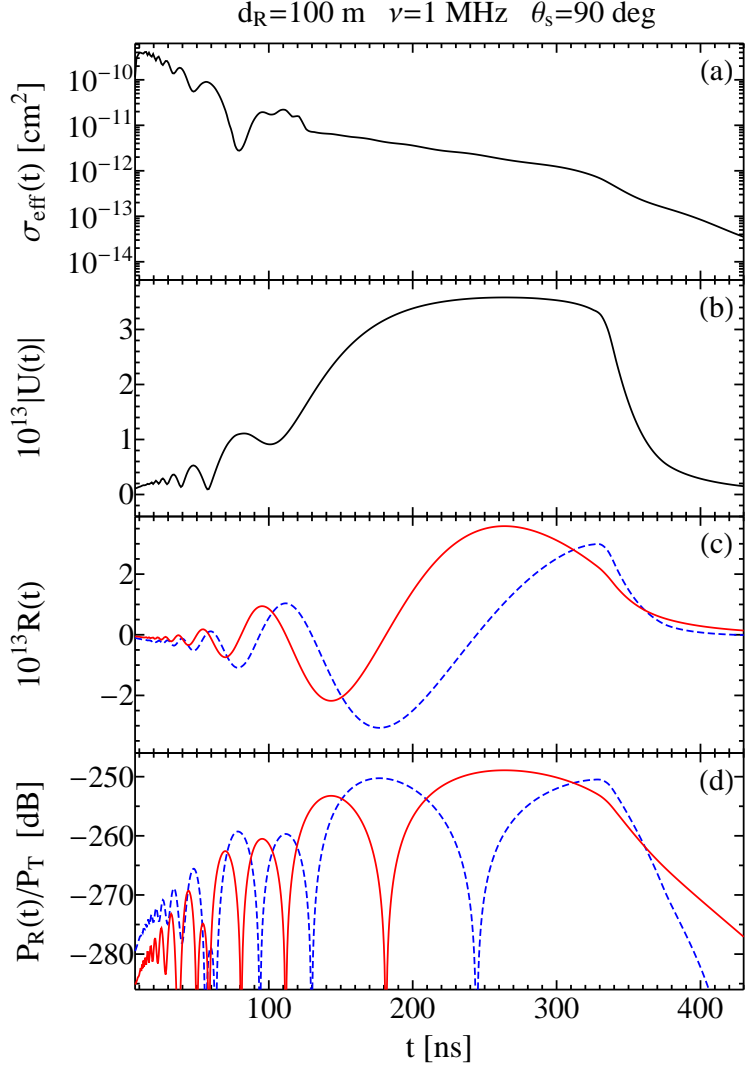}
  \caption{The effective radar cross-section $\sigma_{\rm{eff}}(t)$, the modulus of the complex radar signal at the receiver $|U(t)|$, the waveform of the radar echo $R(t)$, and the ratio of the power received by the detector to the emitted one $P_R(t)/P_T$ calculated for the same shower and radar setup as in Figure \ref{fig:sigma-rL}. The red solid curves represent the waveform and power ratio for such a choice of the initial transmitter phase (phase $\phi_0$ of the emitted signal at $t=0$) that the maximum value of the received power $P_{R,m}$ reaches its largest value $P_{R,\rm{max}}$. Similarly, the blue dotted curves correspond to the smallest $P_{R,m}$ value.}
  \label{fig:U}
 \end{figure}

Figure \ref{fig:U} demonstrates the details of our signal analysis (see Section \ref{SecMax}). It shows the effective radar cross-section $\sigma_{\rm{eff}}$, the modulus of the complex radar signal at the receiver $|U|$, the waveform of the radar echo $R$, and the ratio of the power received by the detector to the emitted one $P_R/P_T$ calculated for the same shower and radar setup as in Figure \ref{fig:sigma-rL}. The time dependence of the power ratio $P_R/P_T$ is a function of the initial transmitter phase (phase $\phi_0$ of the emitted signal at $t=0$). The largest value of $P_R/P_T$, which can be attained at a given time $t$ by varying $\phi_0$ is equal to $|U(t)|^2 $. The red solid curves represent the waveform $R$ and the power ratio $P_R/P_T$ for such a choice of the phase $\phi_0$ that the $\phi_0$-dependent maximum value of the received power $P_{R,m}$ reaches its largest value $P_{R,\rm{max}}$. Similarly, the blue dotted curves correspond to the smallest $P_{R,m}$ value. The difference between $P_{R,m}$ for different choices of the phase $\phi_0$ is usually below a few dB. The local minima in the $P_R(t)/P_T$ functions coincide with the zero points of the corresponding waveforms $R(t)$. This is due to the definition $P_R(t)/P_T = R(t)^2$. Note that $|U|$ and $R$ are shown multiplied by $10^{13}$.

As we can see, the frequencies of the received signal are higher than that of the emitted one, despite the fact that the Thomson scattering preserves the frequency of the scattered radio wave. The observed upshift is caused by the interference of the radio waves reflected from the plasma volume at different stages of its development, i.e. at different times. Thus we observe the expected Doppler effect.

Since the receiver antenna is outside the Cherenkov cone, which is the most common case expected in air shower experiments, the amplitude of the waveform increases with time. This is due to a geometrical effect of the reflections from the lower parts of the atmosphere (the scattering region is closer to the detector). In accordance with the behavior of the signal amplitude, the return power grows with time until the shower front reaches the ground and the plasma has decayed. For cases (not shown here) when the receiver is inside the Cherenkov cone, the time sequence is reversed: the lower part of the shower is seen first and the amplitude decreases with time. 

The enhancement of the radar cross-section at the beginning of the signal is clearly seen in Figure \ref{fig:U}. This is an effect of the increase of the longitudinal extent of the region from which the scattered waves arrive simultaneously to the detector due to the time compression of the received signal. This leads to an increase of the size of the region from which we get the coherent signal. However, the part of the radar echo with the highest frequency upshifts originates from large altitudes. Therefore, despite the fact that the scattered signal is enhanced due to increase of the shower radar cross-section, the return power of the radar echo is small due to the geometrical factor of $r^{-4}$. 

\begin{figure}[bhtps!]
  \centering
    \includegraphics[width=0.75\textwidth]{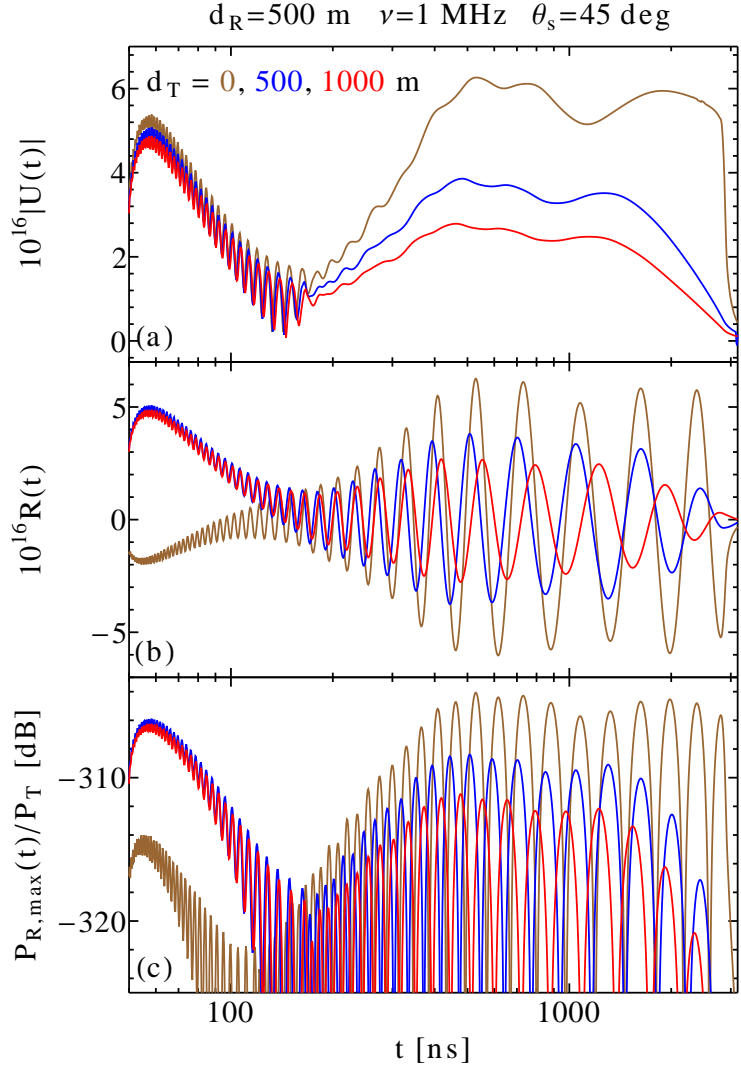}
  \caption{Time dependence of the modulus of the complex radar signal at the receiver $|U(t)|$, the waveform of the radar echo $R(t)$, and the ratio of the power received by the detector to the emitted one $P_{R,\rm{max}}(t)/P_T$ calculated for showers with elevation angle $\theta_s=45^o$, energy $10^{18}$ eV, shower core-receiver distance $d_R=500$ m and core-transmitter distances equal to $d_T=0$ m (brown), 500 m (blue) and 1000 m (red). The frequency of the incident radar wave is $\nu=1$ MHz. See the text for a detailed explanation. 
  }
  \label{fig:45}
 \end{figure}
     \begin{figure}[bhtps!]
  \centering
    \includegraphics[width=0.75\textwidth]{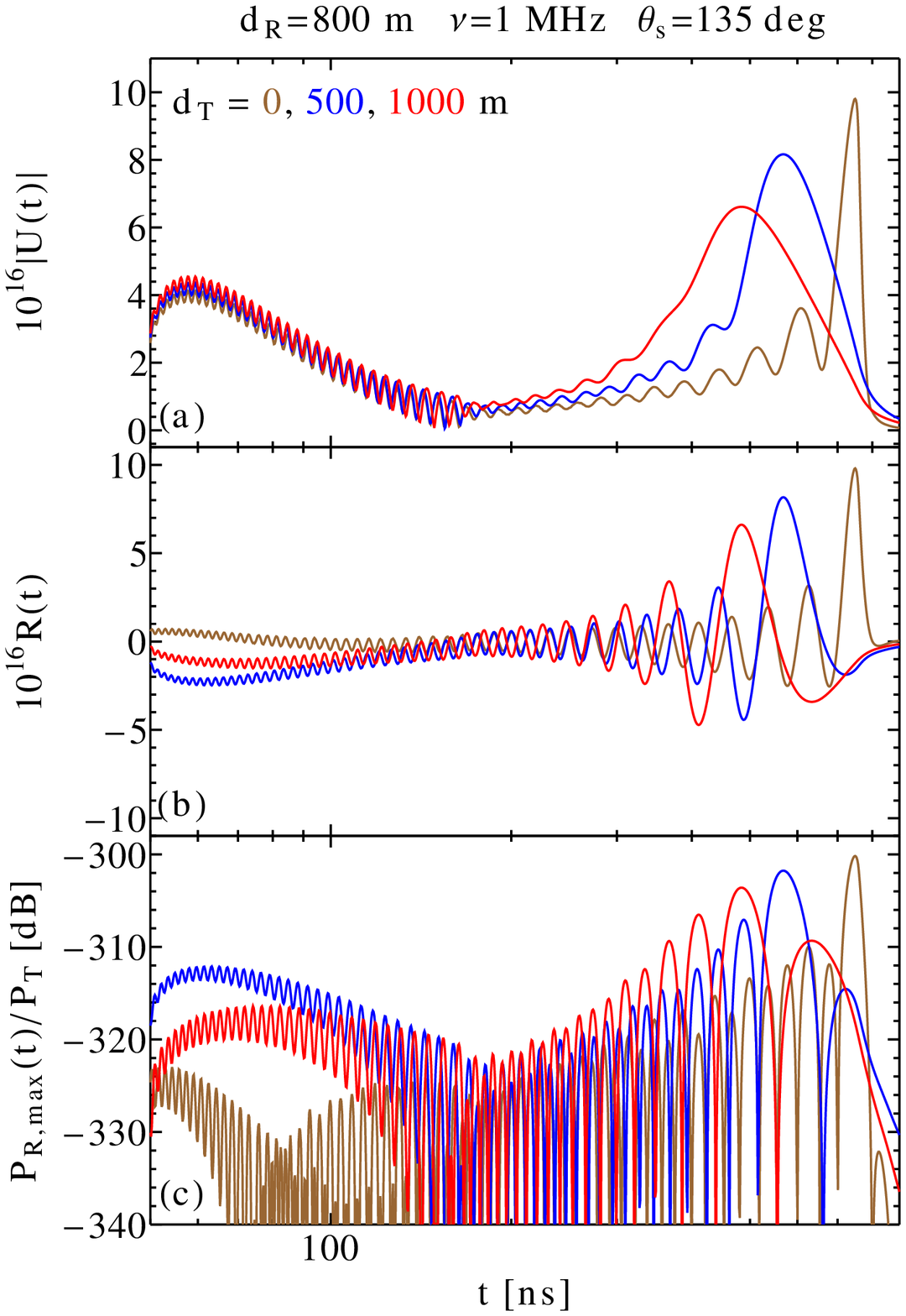}
  \caption{The same as figure \ref{fig:45}, but for the distance between the shower core and the receiver of $d_R=800$ m and elevation angle $\theta_s=135^o$ (the showers approach the receiver).
    }
  \label{fig:135}
 \end{figure}

Now let us consider inclined showers and an incident radio wave with the same frequency of $\nu=1$ MHz.
Figure \ref{fig:45} shows the time dependence of the modulus $|U(t)|$, the ratio of the maximum received power to the emitted one $P_{R,\rm{max}}(t)/P_T$ and the corresponding waveform $R(t)$ for showers with elevation angle $\theta_s=45^{\text{o}}$, energy $10^{18}$ eV, shower core-receiver distance $d_R=500$ m and core-transmitter distances equal to $d_T=0$ m (brown curve), 500 m (blue) and 1000 m (red). The transmitter, receiver and shower axis lie in a common plane perpendicular to the ground. Moreover, the transmitter and receiver antennas are located on the opposite sides of the shower core ($\varphi_T=180^{\text{o}}$, $\varphi_R=0^{\text{o}}$). The shower elevation angle $\theta_s=45^{\text{o}}$ corresponds to the case when the shower passes above the receiver. Note that the entire signal has been artificially shifted to positive times, in order to present the plots in the logarithmic timescale to show the high-frequency, early part of the received signal in more detail.
 
There are three prominent peaks at each of the $|U(t)|$ curves. The first peak corresponds to the time-compressed radar signal reflected at high altitudes. The height of this peak is almost the same for all three considered transmitter positions. This can be explained by the fact that, for the signal produced at high altitude, the distance from the transmitter to the scattering region and to the receiver changes only slightly when varying $d_T$. 
	
The other two peaks represent reflection from lower altitudes, where the change in frequency is moderate and the interplay between the distance to the detector and the frequency upshift (when the shower is approaching the receiver) or downshift (when the shower starts receding from the receiver after passing the point of the closest approach) become important. The heights of these peaks are larger for smaller $d_T$.

The first peak determines the maximum received power $P_{R,\rm{max}}(t)$ only for shower geometries in which the time compression of the reflected signal is high and the signal received from the lower altitudes is sufficiently low, i.e. for inclined showers with the transmitter and receiver antennas far away from the shower core. In this example, this is the case for $d_T=500$ m and $1000$ m.	

Figure \ref{fig:135} shows the same dependences as Figure \ref{fig:45}, but for the shower core-receiver distance of $d_R=800$ m and elevation angle $\theta_s=135^{\text{o}}$ (showers approaching the receiver). For such a geometry, unlike in the previously considered case of the showers passing above the receiver ($\theta_s=45^{\text{o}}$ and $\varphi_R=0^{\text{o}}$), we have only one large peak in $|U(t)|$ at the latest part of the signal. For approaching showers, the frequency of the incoming signal is constantly upshifted and the distance to the receiver decreases with time. This generates a rapid increase in the reflected signal. There is an overall higher time compression of the radar echo, i.e. shorter duration of the signal, when compared to the case of the shower passing above the receiver (as in Fig. \ref{fig:45}). It follows that the peaks in $|U(t)|$ at the latest part of the signals are higher than the corresponding peaks from Figure \ref{fig:45}. This is despite the fact that now the shower core-receiver distance $d_R$ is larger. In contrast, there is only a slight difference in the strength of the signal produced at high altitudes (first $|U(t)|$ peak), because the considered showers (with inclination angles $\theta_s=45^{\text{o}}$ and $\theta_s=135^{\text{o}}$) are seen by the receiver located far away from the scattering region at similar angles. Therefore, the time compression of the received signal is almost identical.   

 \begin{figure}[t]
  \centering
  \includegraphics[width=1\textwidth]{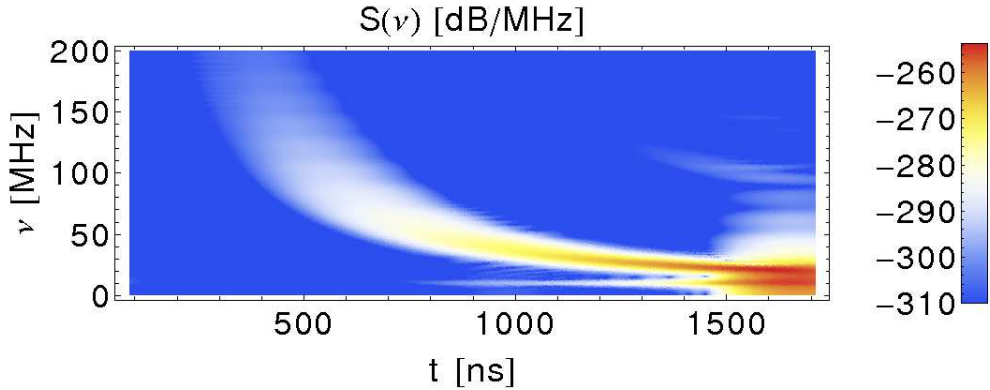}
 \vspace{-0.5cm}
  \caption{Spectrogram of the radar echo of a vertical shower with energy $10^{18}$ eV heading towards the transmitter. The radar frequency and transmitter-receiver distance are equal to $\nu=10$ MHz and $d_R=500$ m, respectively. A 500 ns running time window is used.} 
  \label{fig:sp1}
 \end{figure}
 
In short, our results clearly show that the reflection from the plasma produced by the shower at its maximum does not necessarily generate the maximum of the received power ($P_{R,\rm{max}}$). This is caused by the fact that the distance of the scattering region to the detector and the time compression of the reflected radio wave (frequency upshift) due to the shower geometry are more important than the variation of the number of the plasma electrons with altitude. For vertical showers, the maximum power $P_{R,\rm{max}}$ is usually received from the regions located close to the detector (low altitudes), whereas for the inclined showers that have shower cores located far away from the detector antennas (large $d_T$ and $d_R$), $P_{R,\rm{max}}$ originate from large altitudes not directly related to the maximum of the plasma production by the shower.

Finally, an example of spectrogram of the radar echo  is shown in Figure \ref{fig:sp1}. A vertical shower with energy $10^{18}$ eV heading towards the transmitter ($d_T=0$) is considered here. The radar frequency and transmitter-receiver distance are equal to $\nu=10$ MHz and $d_R=500$ m, respectively. As expected, the frequency decreases with time. Note the low-frequency component at the end of the radar echo, which is caused by the modulation of the received signal by the factor $e^{i\omega t}$ (see equation (\ref{Ucontr2})).	The typical signal consists of two parts: a short signal upshifted to high-frequency with low amplitudes and a long signal with frequency upshifts of only a few and larger amplitudes.

\section{Results and discussion}

\label{diss}

\subsection{Dependence of the maximum received power on the shower energy}

\label{23db}

\begin{figure}[t]
 \vspace{-0.5cm}
  \centering
  \includegraphics[width=0.9\textwidth]{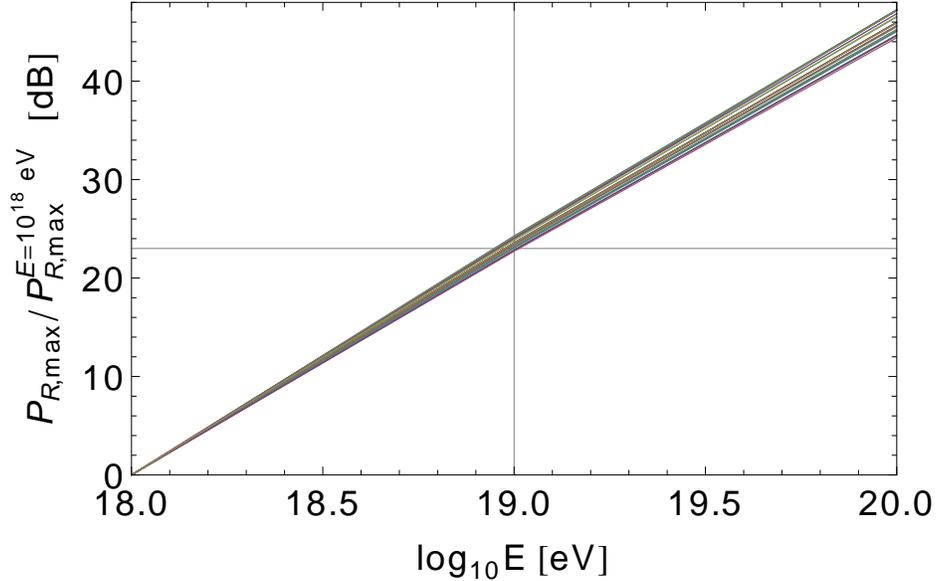}
  \caption{Dependence of
  the maximum received power on shower energy. Each line is derived for different shower geometry and detector setup and shows $P_{R,\rm{max}}$ values scaled to the respective maximum received power for the shower energy of $10^{18}$ eV.
  }
  \label{fig:P-E}
 \end{figure}

Accordingly to the analysis described in Section \ref{SecMax}, we have calculated the maximum received power $P_{R,\rm{max}}$ for each of the simulated radar echoes and prepared a set of plots with the dependences of $P_{R,\rm{max}}$ on the shower energy $E$, shower geometry, and on the radar frequency $\nu$. Figure \ref{fig:P-E} demonstrates the dependence on shower energy. Each line  is derived for a fixed shower geometry and detector setup and shows $P_{R,\rm{max}}$ values scaled to the respective maximum received power for the shower energy of $10^{18}$ eV. The plot is drawn based on the results of the set of simulations performed for several different radar frequencies (1 MHz $\leq \nu \leq$ 50 MHz), for showers with different elevation angles ($45^{\text{o}} \leq \theta_s \leq 135^{\text{o}}$) and different distances of the shower core to the transmitter and the receiver ($d_T$ and $d_R$ up to 1 km).

The dependence of the maximum received power $P_{R,\rm{max}}$ on the shower energy shows a universal scaling. There is about 23 dB increase in power per decade increase in shower energy. This means that for a given shower geometry and detector setup $P_{R,\rm{max}} \sim E^\delta$ with $\delta \approx 2.3$. This scaling is similar to that in coherent scattering, in which the reflected power depends quadratically on the number of scattering particles, in contrast to the linear dependence for incoherent case. The value of the $\delta$ index is an outcome of both the increase in the shower energy deposition and the change of its lateral distribution with increasing shower energy $E$. 

\subsection{Dependence of the maximum received power on the shower geometry}

   \begin{figure}[btps!]
  \centering
  \includegraphics[width=0.9\textwidth]{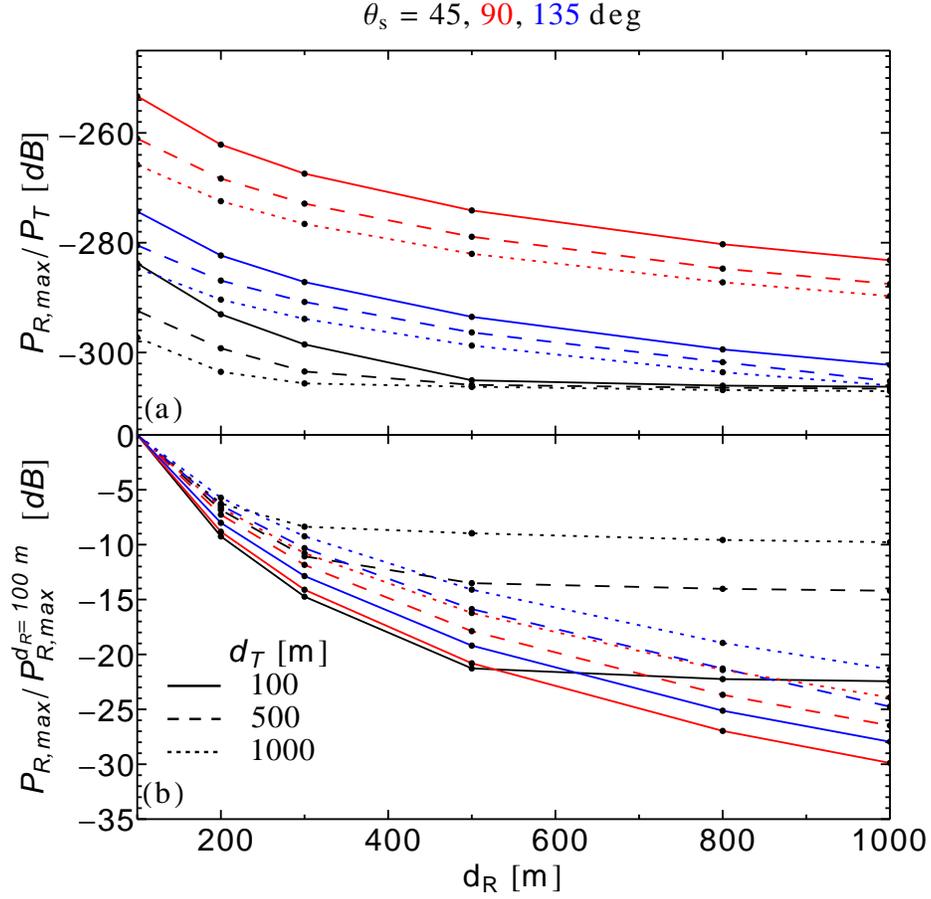}
   \caption{(top panel): Dependence of the ratio of the maximum received power to the emitted one $P_{R,\rm{max}}$/$P_T$ on the shower core-receiver distance $d_R$ and elevation angle $\theta_s$ for showers with energy $10^{18}$ eV. Each line represents a fixed value of the shower core-transmitter distance $d_T$. The transmitter, receiver and shower axis lie in a common plane perpendicular to the ground. The transmitter and receiver antennas are located on the opposite sides of the shower core ($\varphi_T=180^{\text{o}}$, $\varphi_R=0^{\text{o}}$). The frequency of the incident radar wave is $\nu=1$ MHz. (bottom panel): The same as the top panel, but $P_{R,\rm{max}}$ is rescaled to its value at $d_R=100$ m.
  }
  \label{fig:ths}
  \end{figure}

  \begin{figure}[btps!]
  \centering
  \includegraphics[width=0.9\textwidth]{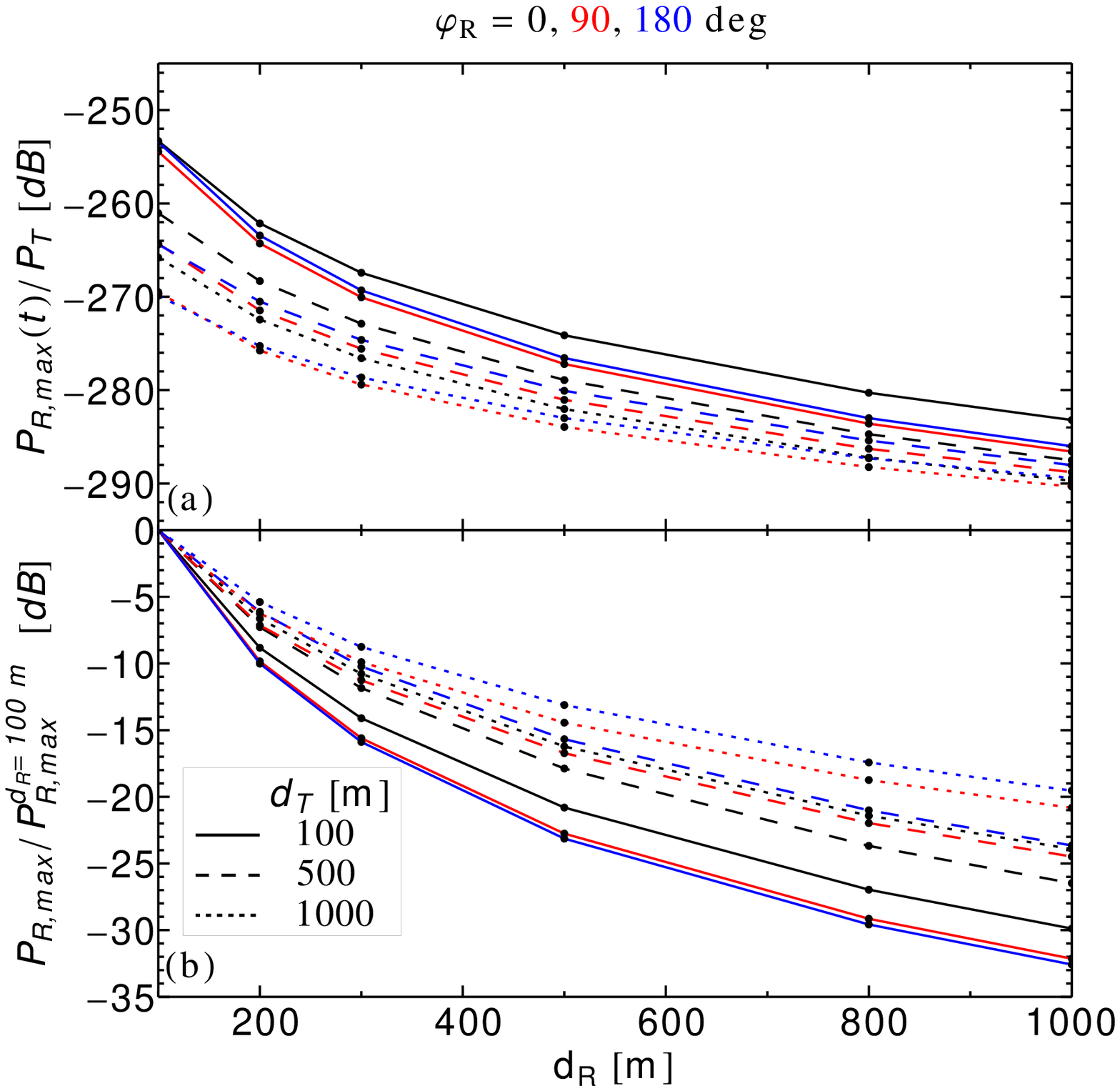}
 \caption{(top panel): Dependence of the ratio of the maximum received power to the emitted one $P_{R,\rm{max}}$/$P_T$ on the shower core-transmitter distance $d_T$ and receiver azimuth angle $\varphi_R$ for vertical showers with energy $10^{18}$ eV. The antennas and the shower core are not collinear; the receiver is placed at various azimuthal angles relative to the transmitter-core line. The radar wave has a frequency of $\nu=1$ MHz. (bottom panel): The same as the top panel, but $P_{R,\rm{max}}$ is rescaled to its value at $d_R=100$ m.}
  \label{fig:phi}
    \end{figure}

The dependence of the maximum received power on the shower geometry and detector setup (varying $\theta_s$, $d_R$, and $\varphi_R$), for showers with energy $10^{18}$ eV and the radar frequency of 1 MHz, is shown in Figures \ref{fig:ths} and \ref{fig:phi}. Figure \ref{fig:ths} shows the dependence of the ratio $P_{R,\rm{max}}$/$P_T$ on the shower core-receiver distance $d_R$ and elevation angle $\theta_s$. Each line represents a fixed value of the shower core-transmitter distance $d_T$. The transmitter, receiver and shower axis lie in a common plane perpendicular to the ground. Moreover, the transmitter and receiver antennas are located on the opposite sides of the shower core ($\varphi_T=180^{\text{o}}$, $\varphi_R=0^{\text{o}}$). 
Note that all of the respective curves should be moved up by about 23 dB and 46 dB for showers with energies $10^{19}$ eV and $10^{20}$ eV, respectively. The bottom panel shows $P_{R,\rm{max}}$ rescaled to its value at $d_R=100$ m. It represents the rate of decrease of the received power with the shower core-receiver distance $d_R$, independent of the actual strength of the signal. The received power $P_{R,\rm{max}}$ decreases by up to about 30 dB when varying $d_R$ from 100 m to 1000 m.

The analogous dependence of the received power on the shower core-transmitter distance $d_T$ (not shown here) is weaker than its dependence on $d_R$. This can be expected as the time compression of the signal (and thus its enhancement) is mostly governed by the relative position of the receiver antenna and the shower core ($d_R$), and the shower inclination angle. The received power decreases by up to about 20 dB when varying $d_T$ from 0 m to 1000 m (up to about 15 dB when varying $d_T$ from 100 m to 1000 m).

The largest power is observed for vertical showers (see Figure \ref{fig:ths}). For $\theta_s=45^{\text{o}}$, the received power is almost constant at large $d_R$. This is due to the fast decrease in the power received from low altitudes as $d_R$ increases. The highly compressed signal arriving from high altitudes (see the first $|U|$ peak in Figure \ref{fig:45}) dominates at large $d_R$. Since it originates in regions located far away from the receiver antenna, its power decreases very slowly with $d_R$. 

The dependence of the maximum received power for vertical showers on $d_R$  and the receiver azimuth angle $\varphi_R$ is shown in Figure \ref{fig:phi} similarly to Figure \ref{fig:ths}. The antennas and the shower core are not collinear; the receiver is placed at various azimuthal angles relative to the transmitter-core line. The signal is strongest for the collinear geometry ($\varphi_R=0^{\text{o}}$).

In general, the dependence of the maximum received power $P_{R,\rm{max}}$ on the shower core-receiver ($d_R$) and the shower core-transmitter ($d_T$) distances are similar for all considered shower geometries. For a fixed geometry (i.e. fixed $\theta_s$, $\varphi_R$, and $\varphi_T$), the maximum power $P_{R,\rm{max}}$ decreases by up to about 30 dB and 15 dB when varying $d_R$ and $d_T$ from 100 m to 1000 m, respectively.
 
\subsection{Dependence of the maximum received power on the radar frequency} 

\label{DepFreq}

\begin{figure}[t]
  \centering
  \includegraphics[width=1\textwidth]{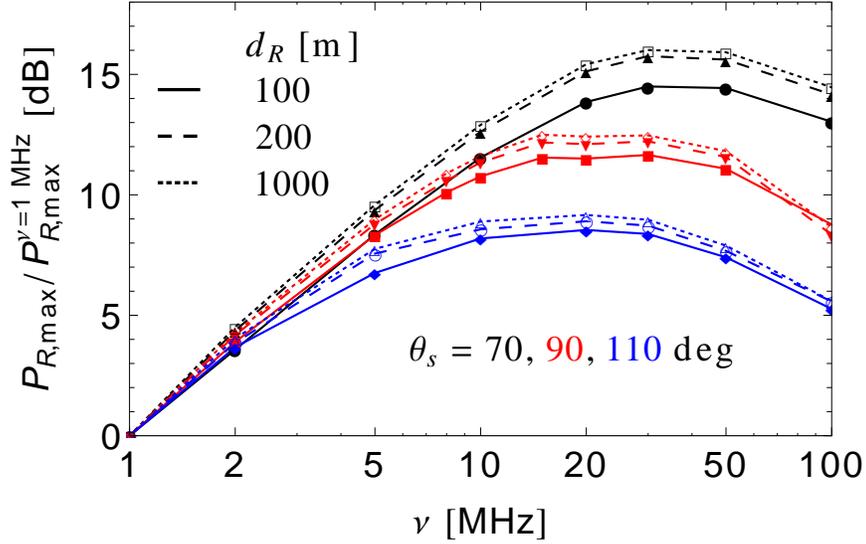}
  \caption{Dependence of the maximum received power on the radar frequency $\nu$ and the shower elevation angle $\theta_s$, scaled to the respective maximum power received for frequency of 1 MHz. Showers heading towards the transmitter are considered. The plane formed by the receiver antenna and shower axis is perpendicular to the ground ($\varphi_R=0 $).
  }
  \label{fig:P-ni}
 \end{figure}

      \begin{figure}[t]
  \centering
    \includegraphics[width=1\textwidth]{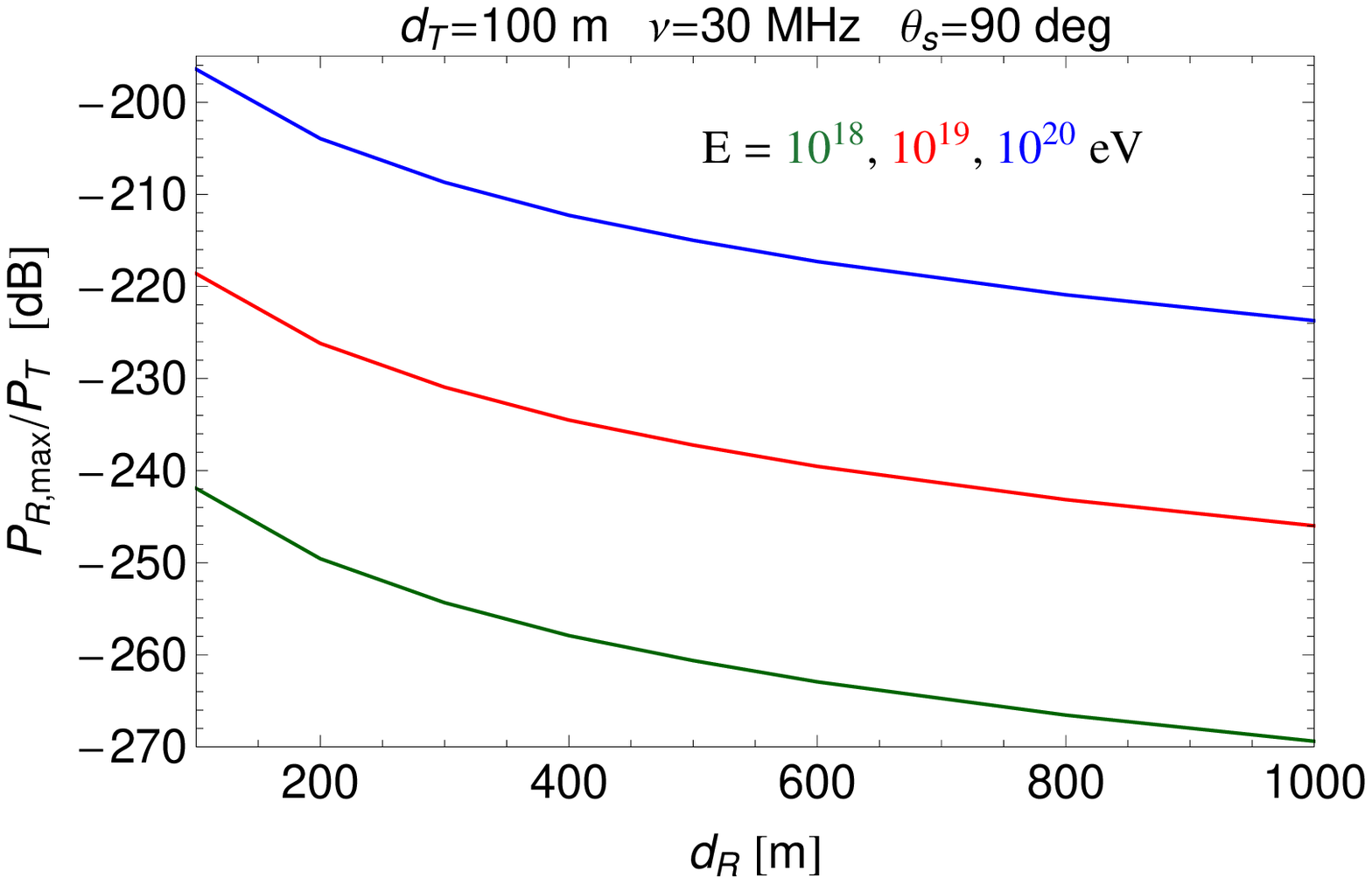}
  \caption{Ratio of the maximum received to the emitted power as a function of the shower core-receiver distance $d_R$ and shower energy $E$, for vertical showers at 100 m from the transmitter. The radar frequency is 30 MHz.}
  \label{fig:Pall}
 \end{figure}

Figure \ref{fig:P-ni} shows the dependence of the maximum received power on the radar frequency $\nu$ and the shower elevation angle $\theta_s$ scaled to the respective maximum power received at the frequency of 1 MHz. Showers heading towards the transmitter are considered. The plane formed by the receiver antenna and shower axis is perpendicular to the ground (with $\varphi_R=0 $).

The size of the region, from which one gets a coherent signal, decreases with decreasing wavelength and destructive interference cancels out the signal from the farther regions of the plasma. At large frequencies, this leads to a decrease in the received power $P_{R,\text{max}}$ with increasing radar frequency $\nu$. On the other hand, $P_{R,\text{max}}$ scales as  $(\omega / \nu_c)^2$ due to the molecular quenching. Therefore, the received power is quenched more for the lower radar frequencies. Effectively, the strongest signal is reached at frequencies around 30 MHz. For vertical showers, the received power increases by about 12 dB when changing the radar frequency from 1 MHz to 30 MHz. This increase is larger for inclined showers. However, this does not translate to the larger values of $P_{R,\rm{max}}$, since in general the inclined showers give weaker signals.

\subsection{Feasibility of radar echo observation}

Figure \ref{fig:Pall} shows the dependence of the received power on the shower core-receiver distance $d_R$ and shower energy $E$, for vertical showers at 100 m from the transmitter. The radar frequency is 30 MHz. The transmitter and receiver antennas are located on the opposite sides of the shower core ($\varphi_T=180^{\text{o}}$, $\varphi_R=0^{\text{o}}$). The values of the power ratio $P_{R,{\rm{max}}}/P_T$, shown in Figure \ref{fig:Pall}, represent the strongest signal that can be attained in the MHz frequency range for $d_T$ and $d_R$ larger than 100 m. The largest value of the power ratio $P_{R,{\rm{max}}}/P_T$ is equal to $-196$ dB for $d_T=d_R=100$ m and for highest energy $E=10^{20}$ eV. The frequency upshift $f_r$ of the received signal at its maximum is small, typically $f_r=2-3$. Thus, the radar echoes of air showers can be observed best at frequencies $\nu_R=60-90$ MHz ($\nu_R$ denotes the frequency of the received signal). Within this frequency range, for a remote location (far away from man-made noise sources) the noise temperature is dominated by the Galactic synchrotron emission and it changes from about $4400$ K to $2000$ K, respectively. It follows that assuming the effective bandwidth of the receiver antenna to be $\Delta \nu_R=10$ MHz, one gets the power of 
the noise $P_n=k_b T_n \Delta \nu_R \geq -95.6$ dBm, where $k_b$ is the Boltzmann constant. Imposing a 5 dB signal-to-noise ratio (SNR) as the detectability threshold leads to the condition $\mathrm{SNR} =P_{R,\rm{max}}/P_n \geq 5$ dB. Assuming that the effective area of the receiver is as large as $A_{R}=10$ m$^2$ and setting the transmitted power to $P_T=10^5$ W, we obtain $P_{R,\rm{max}}=-106$ dBm, i.e. $\rm{SNR}=-10.4$ dB, which is below the detectability threshold. For a grid of $N$ receiving antennas, the signal power scales coherently like $P_{R,\rm{max}} \sim N^2$, whereas the noise is incoherent and scales linearly, i.e. $P_{n} \sim N$. Therefore, we can get a 10 dB increase in the signal-to-noise ratio by utilizing a grid of 10 receiving antennas. Moreover, we estimate that by tilting the antennas to the horizontal position we can get the signal increase from about 8 dB for $d_R=100$ m up to about 24 dB for $d_R=1000$ m. As the result, we obtain $\rm{SNR}=7.6$ dB for $d_R=100$ m, which falls down below 5 dB already around $d_R=200$ m. This demonstrates that in general, detection of the radar echoes of the $10^{20}$ eV showers might be possible, however, the necessity of using high-power transmitters, the large number of receivers and the small spacing of the detector grid (of the order of 300 m) makes the radar technique utilizing the MHz range impractical for air shower detection.

Similarly, we can check the feasibility of detection of the high-frequency part of the radar echo, which arrives from large altitudes and whose strength is a slowly varying function of $d_T$ and $d_R$. For this purpose, the most suitable seems to be the $2-15$ GHz frequency range, which is characterized by very low noise. As the reference value, we can take the maximum power received from an inclined shower with $\theta_s=45^{\text{o}}$. The respective power ratio is $P_{R,\rm{max}}/P_T=-306.5$ dB for a shower with energy $10^{18}$ eV at distances $d_R$ larger than $\sim200$ m (see Figure \ref{fig:ths}). This ratio has to be corrected for the lower values of the collision frequency at high altitudes relative to the value of $\nu_c=4.5$ THz that we have used in simulations. Thus, we increase the received power by 3.5 dB (see Section \ref{Assump}). Moreover, changing the radar frequency from 1 to 30 MHz reduces the molecular quenching by up to 29 dB under assumption that the coherence length does not intervene much. Finally, increasing the shower energy to $10^{20}$ eV gives us another 46 dB increase in the signal strength. Applying these scaling factors, we obtain  $P_{R,\rm{max}}/P_T=-228$ dB.  The typical frequency upshift for the considered case is $f_r\approx 300-500$, which leads to the frequency of the received signal of $\nu_R=9-15$ GHz. Now, if we consider the CROME-like detector setup \cite{bibe:CROME} tuned to the GHz frequency range with the system temperature  $T_n=80$ K, the effective area of the antenna reflector $A_{R}=6.4$ m$^2$ and the effective bandwidth $\Delta \nu_R=600$ MHz, then we get $P_{R,\rm{max}}=-139.9$ dBm, $P_n=-91.8$ dBm and $\rm{SNR}=-48.1$ dB, where as previously we have assumed $P_T=10^5$ W. This clearly demonstrates that the CROME-like detector is unable to see the radar reflections from the plasma produced by air showers. 

To consider possible longer plasma lifetime due to the non-thermal distribution of the electron velocity of the plasma, we have performed simulations with the lifetimes ten times longer than calculated in Section \ref{plasma-lifetime}, i.e. $\tau\approx150-400$ ns within the altitude range of $0-5$ km.  The resulting power of the radar echo is larger by about 17 dB for the radar frequency $\nu=1$ MHz and by about 9 dB for $\nu=10$ MHz, where the second value is lower due to the smaller coherence length. Further increase of the plasma lifetime by another factor of 10 ($\tau=1500-4000$ ns) leads to only a slight increase of the power (by less than 5 dB). We have also checked that applying the proper non-thermal plasma distribution calculated in \cite{bibe:n1} increases the collision frequency by about 16$\%$, which translates to only 1.3 dB decrease in the received power of the radar echo. A fixed value of the collision frequency, which we use in simulations regardless of the altitude, has an impact only on the signal which originates at large altitudes, i.e. on the part of the radar echo with the high frequency upshifts. As was discussed in Section \ref{Assump}, the received power of this signal has to be corrected by +3.5 dB. We estimate the overall accuracy of our results at about 5 dB. This number does not account for a possible inaccuracy in the assumed plasma lifetime and the collision frequency (molecular quenching), which might have a substantial effect.

\subsection{Conclusion}

We present the most complete approach yet to the radar detection of air showers by backscattering. However, the results also apply to a bistatic radar setup, where the transmitter and receiver are separated by a distance comparable to the distances of the detector antennas to the shower. We have shown that, unlike the plasma in a meteor trail, the plasma produced by air showers has to be treated always as underdense. Thus, the reflected radar signal is heavily reduced by molecular quenching. Due to the high temperature of the plasma induced in the air by a shower ($T_e \sim 10^5$ K), the collision frequency $\nu_c$ is estimated at several THz. This value is at least one order of magnitude larger than the commonly used values, which leads to more than two orderds of magnitude increase in molecular quenching of the radar echo. We have summed coherently the radio waves reflected off the individual electrons over the volume of the disk-like ionization trail, taking into account the movement of the wave scattering region behind the relativistically moving shower front. We have performed an extensive set of simulations of radar detection of air showers for different geometries, energies and frequencies of the radar.

Our results show that the radar detection of air showers might be possible only in a special case of showers at small zenith angles with energies $10^{20}$ eV or higher and shower cores situated very close to the transmitter and receiver antennas. It appears therefore that the radar technique of air showers detection on a large scale is impractical due to the necessity to use very high power transmitters and very small spacing of the detector grid. Thus, developing large, efficient and inexpensive air shower detector arrays based on the radar technique is not realistic. 

\section{Note added in proof}
To calculate the effective collision frequency $\nu_c$ in Section 3.3, we apply the initial electron energy distribution, i.e. the distribution immediately after the plasma is created by the passing of the shower front \cite{bibe:n1,bibe:n2}. However, as was recently shown in \cite{samarai}, the energy distribution of the plasma electrons changes very rapidly with time due to the ionization, excitation and attachment processes. After 1 ns all electrons have energies below 1.7 eV. In consequence, the rate of inelastic collisions of the electrons is quickly heavily reduced (from several THz to only a few tens of MHz) and the elastic collisions dominate. This electron spectrum evolution leads only to a moderate decrease of the overall $\nu_c$ values and up to about a 9 dB increase in the received power. Such an increase does not change the conclusion of this paper.

\section{Acknowledgments}
This work has been supported in part by the National Centre for Research and Development (NCBiR) grant ERA-NET-ASPERA/01/11, the ASPERA project BMBF 05A11VKA and the KIT start-up grant 2066995641. We thank O. Deligny and L. Bratek for valuable comments and discussion.

\section{Appendix}

The spatial phase terms $\int_{{\bf r}} n \hspace{0.1cm} {\bf k} \cdot {\rm d} {\bf r}$, $\int_{{\bf r_{sc}}} n \hspace{0.1cm} {\bf k_{sc}} \cdot {\rm d} {\bf r_{sc}}$ and $ \cos \theta$ in equation (\ref{Ucontr}) describing the contribution to the total electric field strength of the radio wave at the receiver antenna are given by
\begin{eqnarray}
	\int_{{\bf r}} n \hspace{0.1cm} {\bf k} \cdot {\rm d} {\bf r} &=& k r n^T_h \rm{,} \label{r-3d} \\
	\int_{{\bf r_{sc}}} n \hspace{0.1cm} {\bf k_{sc}} \cdot {\rm d} {\bf r_{sc}(s)} &=& k r_{\text{sc}} n_h \rm{,}  \label{rsc-3d} \\
		\cos \theta &=& \frac{r^2+r_{sc}^2-d_{TR}^2}{2r_{sc} r}  \label{cos-theta} \rm{,}
\end{eqnarray}
where
\begin{eqnarray}
k &=& |{\bf k}| = |{\bf k_{sc}}| = 2\pi \nu / c \rm{,}  \\
d_{TR} &=& \sqrt{d_T^2+d_R^2 +h_T^2 - 2d_T d_R \cos (\varphi_T-\varphi_R)} \rm{,}  \label{d-form} \\
 n^T_h&=&\frac{1}{(h-h_T)} \int^{h}_{h_T} n(h') dh' \label{nh-redefine} \rm{.}
\end{eqnarray}

  \begin{figure}[t]
  \centering
  \includegraphics[width=0.8\textwidth]{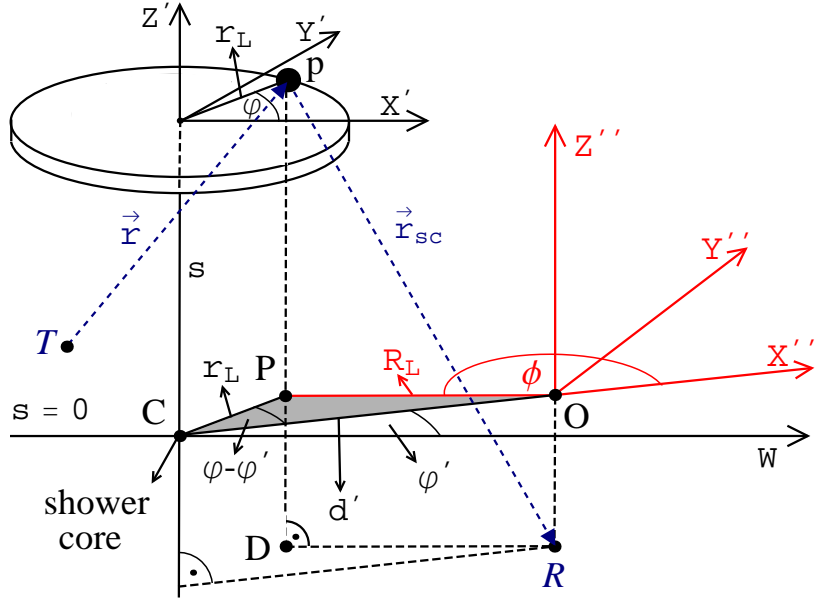}
  \caption{Definition of the cylindrical coordinate system given by the triplet ($R_L$, $\phi$, $s$) used in our calculation (see equation (\ref{total2})). Points O, P and the $W$-axis are the projections of the receiver ($R$), plasma volume element ($p$), and the $X'$-axis on the plane perpendicular to the shower axis ($Z'$) defined by $s=0$. On the same plane lie the $X''$ and $Y''$ axes. The $X''$-axis passes through the shower core ($C$) and $Z''$-axis is parallel to the shower direction and passes through the receiver antenna ($R$). 
}
\label{fig:radar2}
\end{figure}

Here $\theta$, $d_{TR}$, and $\nu$ are the angle between the directions of incident and scattered radio waves, the distance between transmitter and receiver, and the frequency of the emitted radio wave, respectively. To derive formula (\ref{cos-theta})  we have applied the law of cosines to the triangle constructed by the transmitter, receiver, and the plasma volume element ${\rm d}V$. A similar approach was used to derive equation (\ref{d-form}), which expresses the emitter-receiver distance. Note that in order to take into account a non-zero value of the altitude of the receiver antenna $H$, we have to redefine the refractive index in such a way that $n(h)\rightarrow n(h+H)$. Moreover, the altitudes $h$ and $h_T$ should be measured with respect to the receiver level $H$, so that for example the true altitude of the transmitter would be $h_T+H$.

In practice it is more convenient to perform the calculations in the $X''Y''Z''$ coordinate system defined in Figure \ref{fig:radar2}. 
Here, points O, P and the $W$-axis are the projections of the receiver ($R$), plasma volume element (p), and the $X'$-axis on the plane perpendicular to the shower axis ($Z'$) defined by $s=0$. On the same plane lie the $X''$ and $Y''$ axes. The $X''$-axis passes through the shower core ($C$). Moreover, $Z''$-axis is parallel to the shower direction and passes through the receiver antenna ($R$). One can obtain the $X''Y''Z''$ coordinate system from the $X'Y'Z'$ by its  rotation around $Z'$-axis by an angle $\varphi'$ and subsequent translation by $s$. $R_L$ has the meaning of the projection of the distance between the plasma element ($p$) and the receiver ($R$) on the $s=0$ plane.

Note that $V(t)$, which is the contributing volume for a given arrival time t of the signal (see equation (\ref{total}), has a rotational symmetry with respect to the $Z''$-axis, which facilitates defining integration boundaries of the individual contributions to the received signal. In addition, since $n_h$ is a slowly varying function of $h$, we can assume that also $r_{\text{sc}}$ and $s'$ (see equations (\ref{ne-density}), (\ref{sprim}) and (\ref{rsc-eq})) have the same symmetry as $V(t)$, i.e. they are independent of the angle $\Phi$. In contrast, only the electron density $n_e^0(s,r_L)$ is axi-symmetric with respect to the $Z'$-axis. Because of these facts, utilizing the $X''Y''Z''$ frame of reference and the cylindrical system of coordinates ($R_L$, $\phi$, s) (see Figure \ref{fig:radar2}) seems to be the best option.

The integral (\ref{total}) can be rewritten in the following form
\begin{equation}
	U_{\rm{rcv}}(t) =  \iiint\limits_{V(t)} \frac{{\rm d}U_{\rm{rcv}} (t,s,R_L,\phi)}{{\rm d}V} R_L {\rm d}R_L {\rm d}\phi {\rm d}s \rm{,} \label{total2}
\end{equation}
where
\begin{eqnarray}
\frac{{\rm d}U_{\rm{rcv}} (t, s, R_L, \phi )}{{\rm d}V} &=&  U_{T} \sqrt{ \left(\frac{\omega}{\nu_c} \right)^2  \frac{3\sigma_T( 1+\cos^2 \theta ) }{16 \pi}} \frac{  \sqrt{G_{T} \Delta \Omega_{\text{sc}} } }{ r } n_e \nonumber \\
&\times& e^{i(\omega t + \phi_0)} e^{ - ik \left( n_h^T r + n_h r_{\text{sc}} \right)}
 \rm{,} \label{Ucontr2}
\end{eqnarray}
is expressed in terms of $R_L$, $\phi$, and $s$. Formulas for $r_{\text{sc}}$, $r_{L}$, $r$ and $h$ are given by equations (\ref{rsc-eq}), (\ref{rl-eq}), (\ref{r-eq}), (\ref{h-eq}) together with (\ref{coef1})-(\ref{coef2}) and (\ref{xt-eq}), (\ref{xr-eq}), (\ref{dprim}). In the following, we derive these formulas.

The Cartesian coordinates in the $XYZ$, $X'Y'Z'$, and$X''Y''Z''$ frames of reference (see Figures \ref{fig:1} and \ref{fig:radar2}) are given by the triplets ($x$, $y$, $z$), ($x'$, $y'$, $z'$), and ($x''$, $y''$, $z''$). The subscripts $T$, $R$, and $p$ denote the coordinates of the transmitter, receiver, and plasma volume element ${\rm d}V$,  respectively. Therefore, the Cartesian coordinates of the plasma element in the $X'Y'Z'$ frame of reference are
\begin{equation}
\left(
	\begin{array}{c}
x'_p\\
y'_p\\
z'_p
\end{array} \right) = 
	\left(
\begin{array}{c}
 r_L \cos \varphi \\
 r_L \sin \varphi \\
 0
\end{array} \right)  \rm{,}	
\end{equation}
whereas the Cartesian coordinates of the transmitter and receiver in the $XYZ$ frame of reference are 
\begin{eqnarray}
	\left(
	\begin{array}{c}
x_T\\
y_T\\
z_T
\end{array} \right) &=& 
	\left(
\begin{array}{c}
 d_T \cos \varphi_T \\
 d_T \sin \varphi_T \\ 
 h_T
\end{array} \right) \rm{,} \label{xt-eq} \\
	\left(
	\begin{array}{c}
x_R\\
y_R\\
z_R
\end{array} \right) &=& 
	\left(
\begin{array}{c}
 d_R \cos \varphi_R \\
d_R \sin \varphi_R \\
0
\end{array} \right) \label{xr-eq} \rm{.} 
\end{eqnarray}

Transformations between different systems of coordinates are given by
 \begin{equation}
	\left(
	\begin{array}{c}
x'\\
y'\\
z'
\end{array} \right) = 	
	\left(
\begin{array}{ccc}
 \sin \theta_s &0 & -\cos \theta_s  \\
 0 & 1 & 0 \\
 \cos \theta_s &0 & \sin \theta_s  
\end{array} \right) 
\left(
	\begin{array}{c}
x\\
y\\
z
\end{array} \right)
+
	\left(
\begin{array}{c}
 0 \\
 0 \\
 -s
\end{array} \right) \rm{,}
\end{equation}
\begin{equation}
	\left(
	\begin{array}{c}
x''\\
y''\\
z''
\end{array} \right) = 	
	\left(
\begin{array}{ccc}
 \cos \varphi' &\sin \varphi' & 0  \\
  -\sin \varphi' &\cos \varphi' & 0  \\
0 & 0 &1 
\end{array} \right) 
\left(
	\begin{array}{c}
x'\\
y'\\
z'
\end{array} \right)
+
	\left(
\begin{array}{c}
 -d' \\
 0 \\
 s
\end{array} \right) \label{transformation} \rm{.}
\end{equation}

From the above we have
\begin{eqnarray}
	\left(
	\begin{array}{c}
x'_T\\
y'_T\\
z'_T
\end{array} \right) &=& 
	\left(
\begin{array}{c}
x_T \sin \theta_s -z_T \cos\theta_s \\
 y_T \\
 x_T \cos \theta_s +z_T \sin\theta_s -s
\end{array} \right) \rm{,} \\
	\left(
	\begin{array}{c}
x'_R\\
y'_R\\
z'_R
\end{array} \right) &=& 
	\left(
\begin{array}{c}
x_R \sin \theta_s \\
 y_R \\
 x_R \cos \theta_s -s
 \end{array} \right) \rm{.}
\end{eqnarray}

It can be shown that $r_{\text{sc}} = \sqrt{R_L^2 + (z'_R-z'_p)^2}$ (see the $RDp$ triangle in Figure \ref{fig:radar2}) and $d' = \sqrt{{x'_R}^2 + {y'_R}^2}$. Thus, we have
\begin{equation}
	r_{\text{sc}} =  \sqrt{R_L^2 + (s-x_R \cos \theta_s )^2} \rm{,} \label{rsc-eq}
\end{equation}
\begin{equation}
	d' = d_R \sqrt{\cos^2 \varphi_R \sin^2 \theta_s + \sin^2 \varphi_R} \label{dprim} \rm{.}
\end{equation}

The $r^2$ can be expressed by 
\begin{equation}
		r^2 = (x'_p-x'_T)^2+ (y'_p-y'_T)^2 + (z'_p-z'_T)^2 \rm{.}
\end{equation}
Thus we get
\begin{eqnarray}
	r^2 &=& r_L^2 + d_T^2+ h_T^2+s^2-2 r_L  (x_T\sin \theta_s \cos \varphi   +y_T \sin \varphi ) \nonumber \\
	&&-2s x_T  \cos \theta_s   +2 h_T \left( r_L \cos \varphi \cos \theta_s - s \sin \theta_s \right) \label{r2app} \rm{.}
\end{eqnarray}

In order to express $r_L$ in terms of the $X''Y''Z''$ coordinates $R_L$, $\phi$, and $s$ we can use the following formula
\begin{equation}
	r_L =\sqrt{ R_L^2 + d'^2+2R_Ld'\cos \phi }  \rm{,} \label{rl-eq}
\end{equation}
which was obtained by applying the law of cosines to the $POC$ triangle laying at the $s=0$ level.

It is straightforward to show that 
\begin{equation}
\left(
	\begin{array}{c}
x''_p\\
y''_p
\end{array} \right) = 
	\left(
\begin{array}{c}
r_L \cos (\varphi-\varphi')-d' \\
 r_L \sin (\varphi-\varphi') 
 \end{array} \right) 
 =
	\left(
\begin{array}{c}
R_L \cos \phi \\
 R_L \sin \phi 
 \end{array} \right)  
 \rm{,}	\label{app2}
\end{equation}
and
\begin{eqnarray}
	\cos \varphi' &=& \frac{x'_R }{d'} \rm{,} \\ 
	\sin \varphi' &=& \frac{y'_R }{d'} \rm{.}\label{app3}
\end{eqnarray}
Now, using equations (\ref{app2}) - (\ref{app3}) and following expressions 
\begin{eqnarray}
	\cos \varphi &=& \cos(\varphi-\varphi') \cos \varphi' - \sin(\varphi-\varphi') \sin \varphi' \label{cosapp} \\ 
	\sin \varphi &=& \sin(\varphi-\varphi') \cos \varphi' + \cos(\varphi-\varphi') \sin \varphi' 
\end{eqnarray}
we obtain
\begin{eqnarray}
	r_L \cos \varphi &=& \left(d'+ R_L\cos \phi \right)\frac{d_R}{d'} \cos \varphi_R \sin \theta_s  - R_L \sin \phi \frac{d_R}{d'} \sin \varphi_R \rm{,} \label{sub1} \\ 
	r_L \sin \varphi &=& \left(d'+R_L\cos \phi \right) \frac{d_R}{d'} \sin \varphi_R  + R_L \sin \phi \frac{d_R}{d'}\cos \varphi_R \sin \theta_s \label{sub2} \rm{.}
\end{eqnarray}

Substituting equations (\ref{rl-eq}), (\ref{sub1}), and (\ref{sub2}) into equation (\ref{r2app}) one obtains the expression for the distance between the plasma element and the transmitter ($r$). Similarly, substituting equation (\ref{sub1}) into (\ref{hh}), we get a formula for the altitude ($h$). These expressions are the following
\begin{eqnarray}
	r &=& \sqrt{  r_0 + r_1 R_L \cos \phi + r_2 R_L \sin \phi  +r_3 s + R_L^2 +s^2 } \rm{,} \label{r-eq} \\
	h &=& h_0 + h_1 R_L \cos \phi + h_2 R_L\sin \phi  + h_3s  \rm{,}  \label{h-eq} 
\end{eqnarray}
where the coefficients are
\begin{eqnarray}
	r_0 &=& d_T^2+h_T^2 -d'^2 +d' r_1 \rm{,} \label{coef1} \\
	r_1 &=& 2d' -2\sin^2 \theta_s \frac{x_R x_T}{d'}-2\frac{y_R y_T}{d'}+ h_T \sin 2\theta_s \frac{x_R}{d'}  \rm{,} \\
	r_2 &=& 2 \sin \theta_s \frac{y_R x_T- x_R y_T}{d'} - 2 h_T \cos \theta_s \frac{y_R}{d'}   \rm{,}  \\
	r_3 &=& -2x_T \cos \theta_s -2 h_T \sin \theta_s   \rm{,}  \\
	h_0 &=& -\frac{x_R}{2} \sin 2 \theta_s    \rm{,}  \\
	h_1 &=&  \frac{h_0}{d'} \rm{,}  \\
	h_2 &=&  \frac{y_R}{d'}\cos \theta_s \rm{,}  \\
	h_3 &=&  \sin \theta_s \rm{.}  \label{coef2} 
\end{eqnarray}

In the special case of the vertical showers heading towards the transmitter located at the ground level ($\theta_s=\pi/2$, $d_T=0$, $h_T=0$) and the receiver situated at $\varphi_R=0$, equations (\ref{rsc-eq}), (\ref{rl-eq}) , (\ref{r-eq}), (\ref{h-eq}) take the forms
\begin{eqnarray}
r_{\text{sc}} &=& \sqrt{ R_L^2 +s^2 } \rm{,} \\
	r_L &=& \sqrt{R_L^2+d_R^2 + 2 d_R R_L \cos \Phi} \rm{,} \\
	r &=& \sqrt{d_R^2  + 2 d_R R_L \cos \Phi + R_L^2 + s^2} \rm{,} \\
	h &=& s \rm{.}
\end{eqnarray}


\end{document}